\documentclass[twocolumn]{aastex631}

\usepackage{enumitem}
\usepackage{hyperref}
\usepackage{multirow}
\usepackage{soul}
\usepackage{threeparttable}
\usepackage{changepage}
\usepackage{url}
\usepackage{placeins}
\usepackage{rotating}

\shorttitle{UDG}
\shortauthors{Zhou et al.}

\graphicspath{{./}{figures/}}

\begin{document}

\title{An \ion{H}{1} study of a large sample of ultra-diffuse galaxies}

\author[0009-0009-0581-5237]{Ruilei Zhou}
\altaffiliation{E-mail: zhourl@bao.ac.cn}
\affiliation{National Astronomical Observatories, Chinese Academy of Sciences, A20 Datun Road, Beijing 100101, China}
\affiliation{University of Chinese Academy of Sciences, Beijing 100049, China}
\affiliation{CAS Key Laboratory of FAST, National FAST, National Astronomical Observatories, \\
Chinese Academy of Sciences, Beijing 100101, China}
%\email{zhourl@bao.ac.cn}

\correspondingauthor{Ming Zhu}

\author[0000-0001-6083-956X]{Ming Zhu}
\altaffiliation{E-mail: mz@nao.cas.cn}
\affiliation{National Astronomical Observatories, Chinese Academy of Sciences, A20 Datun Road, Beijing 100101, China}
\affiliation{University of Chinese Academy of Sciences, Beijing 100049, China}
\affiliation{CAS Key Laboratory of FAST, National FAST, National Astronomical Observatories, \\
Chinese Academy of Sciences, Beijing 100101, China}
\affiliation{Guizhou Radio Astronomical Observatory, Guizhou University, Guiyang 550000, People's Republic of China}
%\email{mz@nao.cas.cn}

\author{Chuanpeng Zhang}
\affiliation{National Astronomical Observatories, Chinese Academy of Sciences, A20 Datun Road, Beijing 100101, China}
\affiliation{CAS Key Laboratory of FAST, National FAST, National Astronomical Observatories, \\
Chinese Academy of Sciences, Beijing 100101, China}
\affiliation{Guizhou Radio Astronomical Observatory, Guizhou University, Guiyang 550000, People's Republic of China}

\author[0000-0002-4915-4137]{Lin Du}
\affiliation{National Astronomical Observatories, Chinese Academy of Sciences, A20 Datun Road, Beijing 100101, China}
\affiliation{University of Chinese Academy of Sciences, Beijing 100049, China}
\affiliation{Key Laboratory of Optical Astronomy, National Astronomical Observatories, Chinese Academy of Sciences, 20A Datun Road, Chaoyang District, Beijing 100101, China}

\author{Cheng Cheng}
\affiliation{National Astronomical Observatories, Chinese Academy of Sciences, A20 Datun Road, Beijing 100101, China}
\affiliation{Key Laboratory of Optical Astronomy, National Astronomical Observatories, Chinese Academy of Sciences, 20A Datun Road, Chaoyang District, Beijing 100101, China}
\affiliation{Chinese Academy of Sciences South America Center for Astronomy, National Astronomical Observatories, CAS, Beijing 100101, People’s Republic of China}

\author{Qi Guo}
\affiliation{National Astronomical Observatories, Chinese Academy of Sciences, A20 Datun Road, Beijing 100101, China}
\affiliation{University of Chinese Academy of Sciences, Beijing 100049, China}
\affiliation{Institute for Frontiers in Astronomy and Astrophysics, Beijing Normal University, Beijing 102206, People's Republic of China}

\author[0000-0002-1335-6212]{Tian-Wen Cao}
\affiliation{National Astronomical Observatories, Chinese Academy of Sciences, A20 Datun Road, Beijing 100101, China}
\affiliation{Guizhou Radio Astronomical Observatory, Guizhou University, Guiyang 550000, People's Republic of China}
%\email{twcao@bao.ac.cn}

\author{Wenzhe Xi}
\affiliation{University of Chinese Academy of Sciences, Beijing 100049, China}
\affiliation{Yunnan Observatories, Chinese Academy of Sciences, Kunming 650011, China}

\begin{abstract}
By cross-matching the SMUDGes catalog with the FASHI and ALFALFA \ion{H}{1} surveys, we construct an \ion{H}{1}-detected sample of 112 ultra-diffuse galaxies (UDGs) and 48 low-surface-brightness (LSB) galaxies, providing \ion{H}{1}-based redshifts for 76 galaxies for the first time. Using DESI DR1 redshifts, we assemble an \ion{H}{1}-non-detected sample of 168 galaxies for stacking analysis, with detections in two stellar mass bins below $10^{8.5}\,M_\odot$. Combining optical, UV, and \ion{H}{1} data, we investigate stellar masses, star formation rates, gas fractions, and kinematics. We present a systematic analysis of the \ion{H}{1} mass--optical size relation for UDGs and LSBs, showing both populations follow a mass--optical size scaling similar to normal galaxies, suggesting an average \ion{H}{1} surface density, under the assumption that the optical size traces the extent of the \ion{H}{1} distribution. Their stellar mass--size relation indicates nearly constant stellar surface densities for galaxies with central surface brightness $\mu_{0,g} \gtrsim 24\ \mathrm{mag\ arcsec^{-2}}$, independent of UDG or LSB classification. Many UDGs and LSB galaxies are systematically offset from the baryonic Tully--Fisher relation toward higher baryonic masses at a given velocity, consistent with trends found in UDGs with resolved \ion{H}{1} kinematics. Both populations exhibit low star formation efficiencies and long gas depletion times, supporting multiple formation pathways for UDGs.
\end{abstract}


\section{Introduction}
Ultra-diffuse galaxies (UDGs) constitute a population of low-surface-brightness galaxies (LSBs) that are extreme in surface brightness for their large sizes, typically characterized by dwarf-like luminosities combined with unusually large effective radii ($R_e \gtrsim 1.5$ kpc) and very faint central surface brightnesses ($\mu_{0,g} \gtrsim 24$ mag arcsec$^{-2}$).
First identified as a prominent population in the Coma cluster \citep{van2015forty}, UDG candidates have since been reported in large numbers across a wide range of environments, including galaxy clusters (\citealt{lee2020nature}; \citealt{lim2020next}; \citealt{la2022galaxy}), galaxy groups (\citealt{karunakaran2023extending}; \citealt{goto2023systematically}; \citealt{jones2024gas}), and relatively isolated, low-density regions (\citealt{tanoglidis2021shadows}; \citealt{fielder2024all}; \citealt{montes2024almost}). Observations across these environments reveal that UDGs are not a homogeneous class: although many appear to inhabit dark matter halos comparable in mass to those of classical dwarf galaxies, others exhibit more extreme properties, including apparent dark-matter deficiencies \citep{van2018galaxy, van2019second}, unusually high dark-matter fractions (\citealt{beasley2016overmassive}; \citealt{van2016high}), and significant departures from typical galaxy scaling relations \citep{mancera2019off, mancera2020robust}.

The physical origin of UDGs remains debated, with proposed formation pathways invoking both environmental processing and internal evolution.In dense environments, tidal interactions and stripping processes can contribute to making dwarf galaxies more diffuse and/or suppressing their star formation (\citealt{liao2019ultra}; \citealt{sales2020formation}; \citealt{tremmel2020formation}; \citealt{benavides2023origin}; \citealt{2024AJ....168..212F}). In contrast, UDGs found in low-density environments are more commonly interpreted as products of internal mechanisms, such as burst-driven feedback that redistributes baryons to large radii \citep{di2017nihao}, formation within high-spin dark matter halos \citep{amorisco2016ultradiffuse}, or, in some cases, early dwarf–dwarf mergers \citep{wright2021formation}. Distinguishing between these scenarios requires large, uniformly selected samples combined with reliable distance estimates and constraints on gas content.

Although UDGs are commonly regarded as an extreme subset of the broader LSB population, it remains unclear whether UDGs simply represent the high-size or high-spin tail of LSB galaxies or instead constitute a physically distinct class. Both UDGs and LSB galaxies share low stellar surface densities and are often gas-rich, yet growing evidence suggests that UDGs may exhibit systematic differences in their structural scaling relations, kinematic properties, and baryon distributions compared to more typical LSB systems \citep[e.g.,][]{van2016abundance, mancera2019off}. Disentangling the similarities and differences between UDGs and LSB galaxies is therefore essential for assessing whether a single formation pathway can account for the full population of diffuse galaxies or whether multiple mechanisms are required.

Recent wide-area imaging surveys have substantially expanded the census of UDG candidates. The Systematically Measuring Ultra-Diffuse Galaxies (SMUDGes) project (\citealt{zaritsky2023systematically}; Z23 hereafter) constructed a comprehensive catalog of diffuse galaxy candidates using deep optical data from the DESI Legacy Survey \citep{dey2019overview}, which provides homogeneous structural parameters for thousands of objects. While such optical catalogs are essential for identifying UDG candidates and characterizing their morphology, they offer limited insight into the gaseous component that regulates star formation and traces galaxy dynamics. Observations of neutral atomic hydrogen (\ion{H}{1}) are therefore crucial, as \ion{H}{1} both probes the cold gas reservoir and provides kinematic information through its velocity profile. Consequently, \ion{H}{1} measurements offer a direct means of assessing the relative importance of internal processes and environmental effects in shaping UDGs.

Several studies have explored the \ion{H}{1} properties of diffuse galaxies using different UDG selection criteria. \citet{leisman2017almost} identified 115 \ion{H}{1}-bearing ultra-diffuse sources (HUDs) from the ALFALFA survey, including a restrictive subsample of 30 HUDS-R galaxies selected with $r_{\rm eff,g} > 1.5~{\rm kpc}$, $\mu_{0,g} > 24~{\rm mag~arcsec^{-2}}$, and $M_g > -16.8$. \citet{janowiecki2019environment} expanded the ALFALFA-based HUD sample to 252 objects, including 71 HUDS-R galaxies selected using the same restrictive criteria. More recently, \citet{karunakaran2024systematically} conducted targeted GBT \ion{H}{1} observations of 378 optically selected SMUDGes UDG candidates and detected \ion{H}{1} in 110 systems. Based on their physical effective radii and $g$-band central surface brightnesses, they classified the \ion{H}{1} detections into 37 UDGs and 73 LSB dwarfs, adopting $R_{\rm eff} \ge 1.5~{\rm kpc}$ and $\mu_{0,g} \ge 24~{\rm mag~arcsec^{-2}}$ for UDGs. Their LSB dwarf category includes objects that fail either the size or surface-brightness criterion.
\citet{du2024almost} identified a small number of extremely low-mass LSB systems in ALFALFA. Additional constraints have been provided by surveys such as Environmental COntext (ECO) surveys \citep{hutchens2023resolve} and the Mass Assembly of early-Type GaLaxies with their fine Structures (MATLAS; \citealt{poulain2022hi}), though \ion{H}{1}-confirmed UDGs remain relatively rare.

In this work, we construct a large, homogeneous sample of \ion{H}{1}-confirmed UDG candidates by cross-matching the complete SMUDGes catalog \citep{zaritsky2023systematically} with two blind \ion{H}{1} surveys: the ALFALFA survey (\citealt{haynes2018arecibo}) and the FAST All Sky \ion{H}{1} survey (FASHI; \citealt{2024SCPMA..6719511Z, 2026arXiv260631539Z}). ALFALFA provides wide-area \ion{H}{1} coverage over $\sim7000$ deg$^{2}$.
Compared with ALFALFA, FASHI has a higher sensitivity, with a typical rms noise level of $\sim 0.76~{\rm mJy~beam^{-1}}$ at a velocity resolution of $6.4~{\rm km~s^{-1}}$. This higher sensitivity enables FASHI to detect galaxies with \ion{H}{1} masses down to $\sim 10^7~M_\odot$ in the local volume, i.e., at distances of $\lesssim 50~{\rm Mpc}$. Therefore, FASHI allows us to probe significantly lower \ion{H}{1} masses than ALFALFA, extending the census of low-mass, gas-rich galaxies in our UDG sample \citep{2024SCPMA..6719511Z}. The combination of these data sets allows us to assemble one of the largest samples of \ion{H}{1}-confirmed UDGs n properties in a uniform framework, with the aim of constraining their formation and evolutionary pathways.

This paper is organized as follows. Section~2 describes the sample selection and cross-matching procedure. Section~3 presents the derived  galaxies properties such as the stellar masses, star formation rates, and gas-related quantities. Section~4 presents scaling relations and comparisons with previous studies, while Section~5 discusses the implications of our results for UDG formation, with an emphasis on the low-stellar-mass regime. Our conclusions are summarized in Section~6. Throughout this paper, we adopt a flat $\Lambda$CDM cosmology with $H_{0}=70~\mathrm{km~s^{-1}~Mpc^{-1}}$, $\Omega_{\Lambda}=0.7$, and $\Omega_{m}=0.3$.

\section{Data}
\subsection{\ion{H}{1} and optical data}
The FASHI survey \citep{2024SCPMA..6719511Z,2026arXiv260631539Z} is conducted using the 19-beam receiver of FAST, which provides an angular resolution of $2.9^{\prime}$ at 1.4~GHz. Observations are carried out in two modes: the drift-scan mode (\texttt{DecDriftWithAngle}, with a $21.65^{\prime}$ beam separation) optimized for large-area coverage, and a targeted \texttt{MultiBeamOTF} mode used to fill gaps in the mapping region. With a velocity resolution of $6.4,\mathrm{km,s^{-1}}$ and a typical sensitivity of $0.76 \mathrm{mJy beam^{-1}}$, FASHI has so far surveyed $\sim7600~\mathrm{deg}^2$ (about 35\% of the planned footprint) and detected 41,741 extragalactic \ion{H}{1} sources. The current sky coverage mainly lies outside the Arecibo-accessible region, spanning $0^{\mathrm{h}}-17^{\mathrm{h}}30^{\mathrm{m}}$ and $22^{\mathrm{h}}-24^{\mathrm{h}}$, with declinations primarily in the ranges $-6^{\circ}$–$0^{\circ}$ and $30^{\circ}$–$66^{\circ}$.

The ALFALFA survey \citep{haynes2018arecibo} is a blind extragalactic \ion{H}{1} survey conducted with the seven-beam ALFA receiver at Arecibo. In this work, we make use of the complete ALFALFA survey (the $\alpha.100$ release). It covers two main sky regions: a northern area ($7^{\mathrm{h}}30^{\mathrm{m}}$–$16^{\mathrm{h}}30^{\mathrm{m}}$, $0^{\circ}$–$36^{\circ}$) and a southern extension ($22^{\mathrm{h}}$–$03^{\mathrm{h}}$, $0^{\circ}$–$36^{\circ}$), probing distances out to $\sim250$~Mpc. The survey achieves an angular resolution of $3.3^{\prime} \times 3.8^{\prime}$ at 21,cm through a two-pass drift-scan strategy along declination strips.

Optical properties for our galaxies are taken from the complete SMUDGes catalog presented by Z23, which identifies 7,070 UDG candidates from DESI Legacy Survey DR9 imaging over $\sim 20{,}000~\mathrm{deg}^2$. 
From this catalog, we adopt the structural measurements for all objects, along with distance estimates for 1,529 objects and total mass estimates for 1,436 systems.
All sources satisfy a central surface brightness criterion of $\mu_{0,g} > 24~\mathrm{mag~arcsec^{-2}}$.
We note that the optical measurements are based on DESI Legacy Survey DR9 imaging, which may not fully capture the lowest surface-brightness outskirts of some galaxies, and therefore may not trace the complete stellar light distribution in very diffuse systems (e.g. \citealt{pina2024exploring}).

\subsection{Sample selection}
In this subsection, we describe the galaxy sample used in this study. Our sample consists of two complementary subsamples constructed from an optically selected UDG parent sample and then were associated with available \ion{H}{1} survey data. The first subsample includes UDG candidates that are matched with \ion{H}{1} detected sources from available \ion{H}{1} surveys, named the \ion{H}{1}-detected galaxy sample. The second subsample comprises the remaining UDG candidates without individual \ion{H}{1} detections, which are used for a stacking analysis to investigate their average \ion{H}{1} properties. The combined use of these two subsamples allows us to expand the representativeness of our sample and to probe the \ion{H}{1} properties of UDGs across a broader range of gas content. Details of each subsample are presented in the following subsubsections.

\begin{figure*}
\begin{center}
	\includegraphics[width=0.8\textwidth]{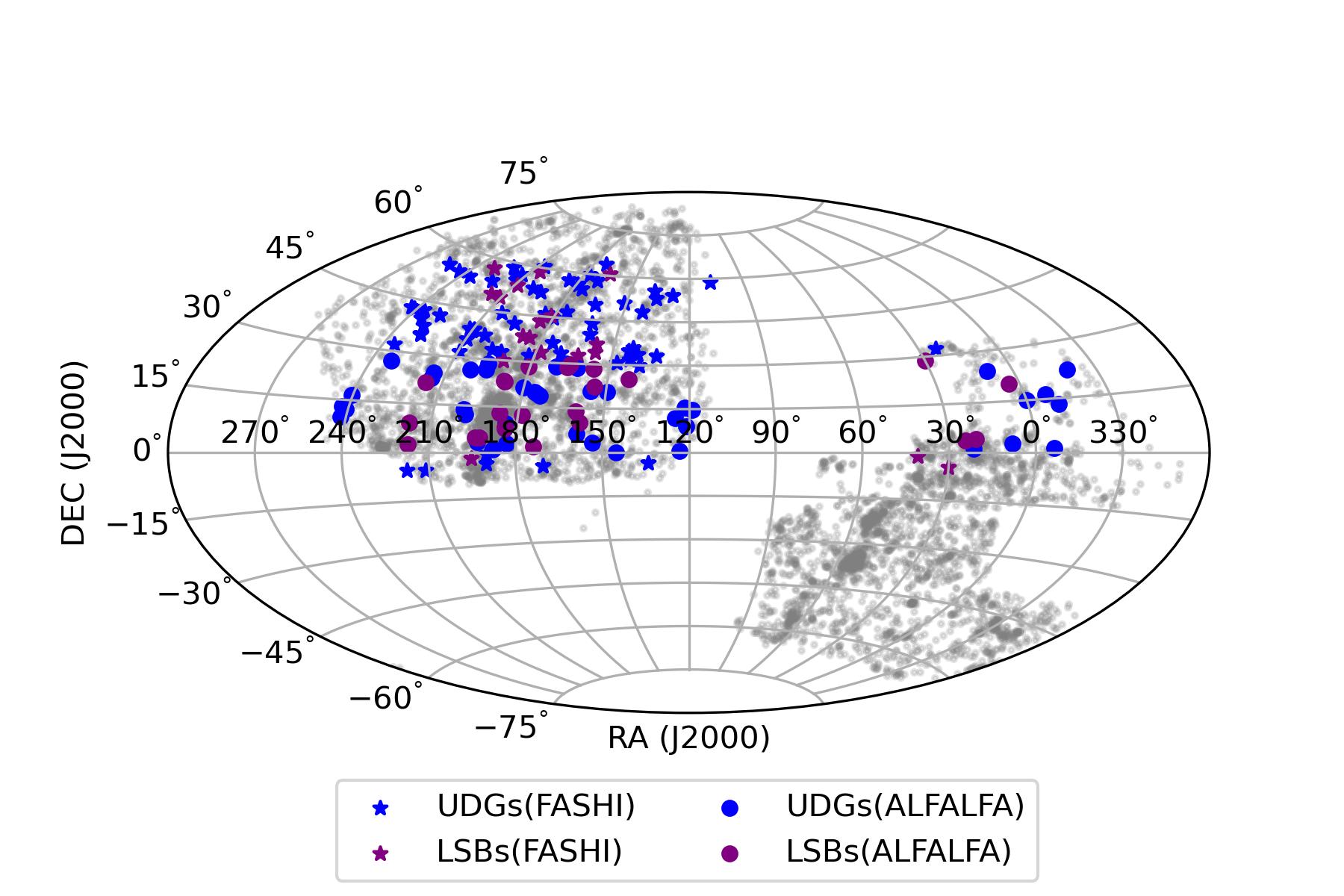}
    \caption{
    The sky distribution of our sample. Blue stars and circles represent UDGs from FASHI and ALFALFA, respectively, while purple stars and circles represent LSBs from FASHI and ALFALFA, respectively. The same symbols are used for these four types of our sample throughout all of the figures, unless otherwise noted.. The complete SMUDGes sample from Z23 is indicated by gray circles. 
    }
\label{fig:sky}
\end{center}
\end{figure*}

\subsubsection{\ion{H}{1}-detected Sample}
This subsample consists of UDG candidates that are successfully matched with \ion{H}{1} detections from existing \ion{H}{1} surveys. It enables us to investigate the galaxy properties in greater detail by incorporating direct \ion{H}{1} observations. The \ion{H}{1}-detected sample was selected following the methodology described below.

Firstly,  we performed cross-matching between the FASHI catalog and the complete SMUDGes catalog from Z23 under the condition $\sigma_{\mathrm{RA}} \leq 1.5^{\prime}$ and $\sigma_{\mathrm{DEC}} \leq 1.5^{\prime}$, as well as between the $\alpha.100$ catalog and the SMUDGes catalog  under the condition $\sigma_{\mathrm{RA}} \leq 1.65^{\prime}$ and $\sigma_{\mathrm{DECs}} \leq 1.65^{\prime}$. The search diameters correspond to the respective resolution of a single beam in FASHI and ALFALFA. The \ion{H}{1} data from FASHI and ALFALFA and the SMUDGes catalog from Z23 provide the essential \ion{H}{1} and optical properties, such as the source position, optical band magnitude, \ion{H}{1} mass, as well as the distance determined by \ion{H}{1} emission. After the cross-matching process, we obtained our parent sample. For the sources matched in both the FASHI and ALFALFA catalogs,  we prioritized to use the FASHI matches.

Secondly, to ensure that the detected \ion{H}{1} emission is genuinely associated with the UDG candidates, we compared the heliocentric velocity of each matched \ion{H}{1} source with the available optical velocity measurements from the SMUDGes catalog. We initially adopted a conservative velocity-difference threshold of $\Delta v \leq 300~{\rm km~s^{-1}}$ as a broad pre-selection criterion to avoid prematurely excluding possible associations, given the heterogeneous availability and quality of optical velocities and the possibility of confusion within the single-dish \ion{H}{1} beam. We emphasize that this threshold was used only for the initial screening rather than as the final acceptance criterion.
We find that the vast majority of matched sources have velocity differences well below $50~\mathrm{km~s^{-1}}$. A small number of UDG candidates show larger offsets; in particular, two sources with $\sigma_{\mathrm{velo}} > 100~\mathrm{km~s^{-1}}$ are likely mismatches and are therefore excluded from the final sample. A few additional sources have intermediate offsets ($50$–$75~\mathrm{km~s^{-1}}$), but their properties are consistent with the rest of the sample and are retained.\ We then visually inspected each remaining candidate using the DESI Legacy Imaging Survey Sky Viewer\footnote{\url{http://legacysurvey.org/viewer}}, examining regions of radius $3'$ (FASHI) or $3.3'$ (ALFALFA) centered on the \ion{H}{1} centroid. Any candidate whose \ion{H}{1} emission was likely contaminated by nearby galaxies was excluded. In this way, we removed 122 sources and obtained 202 \ion{H}{1} sources as the counterparts of the UDG candidates.

Thirdly, we compared our results with previous \ion{H}{1} studies metioned in Section 1 (K24; \citealt{leisman2017almost}; \citealt{du2024almost}; \citealt{hutchens2023resolve}; \citealt{poulain2022hi}). All previously \ion{H}{1}-studied sources for the SMUDGes are presented in Table~\ref{table:previous}, and they were excluded from our final sample. The source SMDG0134058+320619 was in the sample of K24, but not detected by GBT. It was successfully detected by FAST, thus we retained it in our final sample. Both SMDG1031105+343018 and SMDG1031120+343013 were associated with the same \ion{H}{1} detection, and they were visually confirmed to be the same galaxy, thus we retained SMDG1031105+343018  in our final sample.

\begin{table}
    \centering
    \caption{SMUDG candidates from previous \ion{H}{1} studies} 
     \label{table:previous} 
    \begin{tabular}{cc}
        \toprule % 顶部粗横线
        Name & References \\
        (1) & (2) \\
        \hline % 中间横线
        SMDG1411177+481142 & P22 \\
        SMDG1418561+362900 & P22 \\
        SMDG1220187+280132 & P22 \\
        SMDG1241191+013254 & P22 \\
        SMDG1057274+091029 & P22 \\
        SMDG0015213+010428 & L17 \\
        SMDG0149384+304050 & L17 \\
        SMDG0854108+285805 & L17 \\
        \hline % 底部粗横线
    \end{tabular}
    \begin{tablenotes}\footnotesize
 \item \textbf{NOTE.} The first 8 rows of this table are shown here. The full table is available online in machine - readable format. Column definitions: 
 
        (1) Adopted SMUDGes UDG candidate name;
        
        (2) Previous references containing SMUDG candidates. P22=\citep{poulain2022hi}, L17=\citep{leisman2017almost}, H23=\citep{hutchens2023resolve}, K24=\citep{karunakaran2024systematically}.
\end{tablenotes}
\end{table}

Our final \ion{H}{1}-detected sample comprised 160 UDG candidates with \ion{H}{1} detections. We calculated the physical effective radius ($R_{\mathrm{eff}}$) for each UDG candidate, using the angular effective radius from the complete SMUDGes catalog and distances provided by \ion{H}{1} emission. All galaxies are drawn from the SMUDGes catalogue of Zaritsky et al. (2023), in which sources already satisfy the low surface brightness criterion $\mu_{0,g} > 24~\mathrm{mag~arcsec^{-2}}$. UDGs in that work are defined using both surface brightness and size criteria, specifically $\mu_{0,g} > 24~\mathrm{mag~arcsec^{-2}}$ and $R_{\mathrm{eff}} > 1.5~\mathrm{kpc}$. We adopt this definition directly.
In this work, we therefore classify galaxies based on their effective radius into UDGs ($R_{\mathrm{eff}} > 1.5~\mathrm{kpc}$) and lower-mass LSB galaxies ($R_{\mathrm{eff}} < 1.5~\mathrm{kpc}$), noting that all sources already satisfy the surface brightness requirement. 
Here, UDGs represent a subset of LSB galaxies that satisfy the UDG selection criteria, while the term ``LSB galaxies'' in this work specifically refers to LSB systems that do not meet the UDG definition, unless stated otherwise.
Table~\ref{table:udg_lsb} presents the properties of our sources, with the last column indicating their subtypes, namely UDG--FASHI, UDG--ALFALFA, LSB--FASHI, and LSB--ALFALFA.
We initially separated the ALFALFA and FASHI detected samples to examine potential systematic differences arising from their different survey sensitivities and selection functions. Based on the subsequent analysis presented in this work, we found that the two samples are consistent with each other in all relevant properties. Nevertheless, we retained this distinction in the following figures to indicate the original survey from which each galaxy is drawn.
We thus divide our \ion{H}{1}-detected sample into 112 UDGs and 48 LSB galaxies, all of which are newly confirmed.
To the best of our knowledge, this is one of the largest homogeneous samples of H I-confirmed UDGs constructed from SMUDGes under a uniform surface-brightness and physical-size definition, broadly comparable to the restrictive HUDS-R size and surface-brightness criteria used in previous ALFALFA-based studies (e.g. \citealt{leisman2017almost}; \citealt{janowiecki2019environment}).
Figure~\ref{fig:sky} shows the sky distribution of our sample.

We note that, as our sample is constructed from an optically selected UDG catalogue, it may be subject to selection biases against the most diffuse, gas-rich low-surface-brightness systems that are difficult to detect in shallow optical imaging.

By cross-matching our sample with DESI Data Release~1 \citep{abdul2025data}, we identified an additional 11 UDG candidates with optical velocities consistent with their \ion{H}{1} counterparts, including four previously cataloged with optical velocities in the complete SMUDGes catalog from Z23. Notably, 86 UDG candidates in our sample exhibited matched optical velocities and heliocentric velocities from \ion{H}{1} emission, robustly confirming the association between them and corresponding \ion{H}{1} detections. 
For the remaining 76 sources, redshifts are reported for the first time based on their \ion{H}{1} emission, among which 61 are provided by FASHI and 15 by ALFALFA.

We calculated the detection rates of UDGs and LSBs. The complete SMUDGes catalog from Z23 contains 1622 sources within the FASHI survey footprint and 3018 sources within the ALFALFA survey footprint. Therefore, the detection rate in the FASHI sky area is 5.36\%, while that in the ALFALFA sky area is 2.92\% (including 13 sources overlapping between FASHI and ALFALFA). The higher detection rate in the FASHI area can be attributed to the higher sensitivity of FAST. However, during our data processing, to ensure the association between the \ion{H}{1} detections and their optical counterparts, we removed sources that might be affected by confusion. As a result, the detection rates we report should be considered lower limits. Nevertheless, even these rates remain quite low, indicating that we can only detect \ion{H}{1}-rich sources. This suggests that the vast majority of UDGs and LSBs are in fact \ion{H}{1}-poor.

\subsubsection{\ion{H}{1}-non-detected Sample for Stacking Analysis}
Although the sensitivities of FASHI and ALFALFA are not sufficient to detect \ion{H}{1} in every individual UDG candidate, their large sky coverage allows a statistical investigation of the overall \ion{H}{1} properties of UDG candidates without \ion{H}{1} detections. Combined with the abundant spectroscopic redshift samples from DESI, this provides an ideal opportunity to study this problem through stacking analysis. By cross-matching with the DESI DR1 \citep{abdul2025data}, together with the redshifts already provided in the complete SMUDGes catalog of Z23, we constructed a galaxy sample of 168 UDG candidates with spectroscopic redshifts, all located within the FASHI and ALFALFA footprints but without any \ion{H}{1} detections. The stellar mass estimation is described in Section~3.1 and we divided the sample into four stellar mass bins: $6.0 < \log(M_\ast/M_\odot) < 8.0$, $8.0 < \log(M_\ast/M_\odot) < 8.5$, $8.5 < \log(M_\ast/M_\odot) < 9.0$, and $9.0 < \log(M_\ast/M_\odot) < 9.6$. To avoid confusion effects caused by multiple galaxies within a single telescope beam, we examined our galaxy sample and confirmed that there are no sources within one beam radius of FASHI and ALFALFA with velocity differences smaller than 300~km~s$^{-1}$. 
Such confusion would otherwise result in nearly identical \ion{H}{1} spectra extracted at different positions and lead to an overestimation of the average \ion{H}{1} mass.

To obtain the average \ion{H}{1} flux of the galaxy samples, we extracted the \ion{H}{1} spectra at the coordinates of each galaxy and stacked these spectra using the HISS \citep{healy2019hiss} code\footnote{\url{https://github.com/healytwin1/HISS}}. We assigned a weight of $1/\mathrm{rms}^2$ to each spectrum, where rms denotes the root-mean-square noise. The aperture diameter for the \ion{H}{1} spectral extraction was set to $6\arcmin$, which is sufficient to cover the entire galaxy. We employed a stacking procedure based solely on the extracted \ion{H}{1} spectra. The stacking results were presented in Section 4.4.

\subsection{UV data}
To trace recent star formation, we use archival ultraviolet data from the Galaxy Evolution Explorer (GALEX; \citealt{2005ApJ...619L...1M}). GALEX provides imaging in two bands, the near-ultraviolet (NUV; 1770--2730\,\AA) and the far-ultraviolet (FUV; 1350--1780\,\AA), which are sensitive to star formation over the past $\sim$100 Myr. UV photometry is measured for the UDGs and LSBs where available. Photometric accuracy beyond 1.1$^{\circ}$ from the field center is severely affected by edge artifacts \citep{bianchi2011catalogues}.

Through a search on the Mikulski Archive for Space Telescopes (MAST), we found 141 sources within the coverage of GALEX, and we confirmed that they were located more than 3$^{\prime}$ away from the edge of the field of view to ensure photometric accuracy after checking the UV images. The UV data were accessed via \texttt{gPhoton} \citep{2016ApJ...833..292M}, which is a database product with associated software packages providing a user-friendly way to obtain and process photon-level images from the GALEX General Release (GR) 6/7. We downloaded the intensity maps for each GALEX frame listed in column (26) of Table~\ref{table:udg_lsb} using the \texttt{gMap} module of \texttt{gPhoton}.

\section{Analysis} 
In this section, we derive the stellar masses and star formation properties of the galaxies based on the optical and UV data. It should be noted that for sources in our sample, if any optical measurement from Z23 has a flag=1, this indicates that the source is extrapolated and does not participate in the subsequent calculations or discussions. These flags are indicated in Table~\ref{table:udg_lsb}.

\subsection{Stellar Mass}
Stellar mass ($M_\ast$) is a fundamental parameter in galaxy studies and strongly correlates with a wide range of observable properties, such as color, morphology, and chemical enrichment (\citealt{blanton2005relationship,mouhcine2007environmental,brough2013galaxy}).
Given the lack of broad multi-band photometry for our sample, we estimate stellar masses using color–mass-to-light ratio relations (CMLRs), which relate a galaxy’s optical color and single-band luminosity to $M_\ast$. Such relations are commonly calibrated based on stellar population synthesis models or SED-based measurements, and have been established for a variety of galaxy populations, including spiral galaxies \citep{bell2003optical}, irregular dwarfs \citep{herrmann2016mass}, and low-surface-brightness galaxies \citep{du2020stellar}.

Given the absence of multi-band photometry for our sample, we derived stellar masses using the $g$-band magnitude ($m_g$) and $g-r$ color listed in Table~\ref{table:udg_lsb}, adopting the CMLR prescriptions of \citet{zhang2017impact}, which are based on a lognormal \citet{chabrier2003galactic} stellar initial mass function (IMF). 
This choice is consistent with recent studies of similar galaxy populations (e.g. \citealt{karunakaran2024systematically}) and provides a widely used calibration applicable to low-surface-brightness systems. To assess the impact of the adopted CMLR, we recompute stellar masses using the relation of \citet{herrmann2016mass}, which is specifically calibrated for irregular dwarf galaxies. We find that the resulting stellar masses differ by less than 0.1 dex for all sources, indicating that the choice of CMLR introduces only minor systematic differences. This is significantly smaller than the typical uncertainties in our stellar mass estimates ($\gtrsim 0.3$ dex), and therefore does not affect our conclusions.
All optical magnitudes were corrected for Galactic extinction. No correction for intrinsic dust extinction was applied, given the expected low dust content of UDGs and LSB galaxies. $K$-corrections were neglected due to the low redshifts of our galaxies. Because several photometric quantities in the SMUDGes catalog exhibit asymmetric uncertainties (see Z23), we propagated errors using the \texttt{asymmetric\_uncertainty} Python package \citep{gobat2022asymmetric}, which follows the formalism of \citet{starling2008gamma}. We adopt this procedure for all derived parameters throughout this work, including the \ion{H}{1}-to-stellar mass ratio ($M_{\rm HI}/M_\ast$) and the physical effective radius ($R_{\rm eff}$).

\newpage

\begin{adjustwidth}{0cm}{-15cm}
\begin{sidewaystable*}
\centering
\hspace*{-5cm}% 向左移动 2 cm
\resizebox{1.2\textwidth}{!}{%
    \begin{threeparttable}
    \vspace{-280pt} % 与参考格式保持一致的垂直偏移
    \caption{Properties of UDG and LSBs}
    \label{table:udg_lsb}

    % 第一部分表格：光学与基础参数
    %\setlength{\tabcolsep}{2pt}
    
    \begin{tabular}{ccccccccccccccc}
    \toprule
    Name & Type & FASHI\_ID/AGC & RA$_\mathrm{OC}$ & DEC$_\mathrm{OC}$ & RA$_\mathrm{HI}$ & DEC$_\mathrm{HI}$ & $m_g$ & $m_g$\_flag & $m_r$ & $m_r$\_flag & $\mu_{0,g}$ & $\mu_{0,g}$\_flag & $r_\mathrm{eff}$ & $r_\mathrm{eff}$\_flag \\
    & & & deg & deg & deg & deg & mag & & mag & & mag/arcsec² & & arcsec & \\
    (1) & (2) & (3) & (4) & (5) & (6) & (7) & (8) & (9) & (10) & (11) & (12) & (13) & (14) & (15) \\ 
    \hline
    SMDG0134058+320619 & UDG-FASHI & 20230051983 & 23.52 & 32.11 & 23.53 & 32.11 & $19.25^{+0.07}_{-0.15}$ & 0 & $18.86^{+0.07}_{-0.15}$ & 0 & $24.00^{+0.07}_{-0.15}$ & 0 & $7.73^{+0.31}_{-0.29}$ & 0 \\
    SMDG0711577+584016 & UDG-FASHI & 20230039485 & 107.99 & 58.67 & 107.98 & 58.67 & $17.70^{+0.12}_{-0.16}$ & 0 & $17.29^{+0.15}_{-0.38}$ & 0 & $24.07^{+0.06}_{-0.14}$ & 0 & $10.60^{+2.50}_{-0.81}$ & 0 \\
    SMDG0832245+541026 & UDG-FASHI & 20230035867 & 128.1 & 54.17 & 128.09 & 54.18 & $18.66^{+0.09}_{-0.16}$ & 0 & $18.18^{+0.10}_{-0.14}$ & 0 & $24.07^{+0.08}_{-0.11}$ & 0 & $8.37^{+1.06}_{-0.52}$ & 0 \\
    SMDG0850389+331004 & UDG-FASHI & 20230009975 & 132.66 & 33.17 & 132.67 & 33.18 & $19.51^{+0.05}_{-0.12}$ & 0 & $19.27^{+0.07}_{-0.14}$ & 0 & $24.32^{+0.06}_{-0.14}$ & 0 & $5.63^{+0.15}_{-0.11}$ & 0 \\
    SMDG0856308-033623 & UDG-FASHI & 20230051895 & 134.13 & -3.61 & 134.12 & -3.6 & $19.41^{+0.05}_{-0.13}$ & 0 & $19.28^{+0.07}_{-0.15}$ & 0 & $24.08^{+0.05}_{-0.15}$ & 0 & $5.39^{+0.13}_{-0.16}$ & 0 \\
    SMDG0905069+530547 & UDG-FASHI & 20230034629 & 136.28 & 53.1 & 136.28 & 53.09 & $18.89^{+0.08}_{-0.11}$ & 0 & $18.60^{+0.09}_{-0.15}$ & 0 & $24.43^{+0.07}_{-0.09}$ & 0 & $8.36^{+0.76}_{-0.40}$ & 0 \\
    SMDG0910265+553452 & UDG-FASHI & 20230037344 & 137.61 & 55.58 & 137.6 & 55.6 & $19.41^{+0.11}_{-0.19}$ & 0 & $19.04^{+0.14}_{-0.21}$ & 0 & $24.06^{+0.13}_{-0.18}$ & 0 & $6.18^{+1.17}_{-0.46}$ & 0 \\
    SMDG0915558+295527 & UDG-FASHI & 20230060440 & 138.98 & 29.92 & 138.96 & 29.92 & $18.87^{+0.05}_{-0.14}$ & 0 & $18.66^{+0.07}_{-0.14}$ & 0 & $24.13^{+0.05}_{-0.14}$ & 0 & $9.09^{+0.26}_{-0.25}$ & 0 \\
    \hline
    \end{tabular}
    %\end{adjustwidth}
    
    \vspace{12pt}
    
    % 第二部分表格：动力学与物理参数
    %\setlength{\tabcolsep}{2pt}
    
    \begin{tabular}{ccccccccccccc}
    \toprule
    Name & b/a & b/a\_flag & $W_{20}$ & $W_{50}$ & V$_\mathrm{helio}$ & Dist & $\log M_\mathrm{HI}$ & $\log M_\ast$ & $R_\mathrm{eff}$ & $m_\mathrm{NUV}$ & GALEX\_Frame & $Z_\mathrm{DESI}$ \\
    & & & km/s & km/s & km/s & Mpc & log[M$_\odot$] & log[M$_\odot$] & kpc & mag & & \\
    (1) & (16) & (17) & (18) & (19) & (20) & (21) & (22) & (23) & (24) & (25) & (26) & (27) \\ 
    \hline
    SMDG0134058+320619 & $0.45^{+0.03}_{-0.02}$ & 0 & $134.22\pm10.93$ & $67.69\pm7.29$ & $6393.01\pm3.64$ & $88.57\pm4.43$ & $8.90\pm0.08$ & $8.14^{+0.26}_{-0.28}$ & $3.32^{+0.21}_{-0.21}$ & $20.94\pm0.10$ & AIS\_63\_1\_17 &  \\
    SMDG0711577+584016 & $0.94^{+0.02}_{-0.02}$ & 0 & $48.69\pm3.68$ & $29.89\pm2.45$ & $3360.72\pm1.23$ & $43.67\pm2.18$ & $8.73\pm0.08$ & $8.17^{+0.63}_{-0.61}$ & $2.24^{+0.54}_{-0.20}$ & $19.82\pm0.12$ & AIS\_75\_1\_30 & 0.01 \\
    SMDG0832245+541026 & $0.79^{+0.03}_{-0.03}$ & 0 & $89.71\pm4.72$ & $58.49\pm3.15$ & $9568.31\pm1.57$ & $134.60\pm6.73$ & $9.29\pm0.07$ & $8.87^{+0.27}_{-0.31}$ & $5.46^{+0.74}_{-0.44}$ & / & / &  \\
    SMDG0850389+331004 & $0.56^{+0.03}_{-0.02}$ & 0 & $97.35\pm5.15$ & $59.49\pm3.43$ & $8093.58\pm1.72$ & $115.80\pm5.79$ & $8.80\pm0.06$ & $8.03^{+0.24}_{-0.23}$ & $3.16^{+0.18}_{-0.17}$ & / & AIS\_215\_1\_29 & 0.03 \\
    SMDG0856308-033623 & $0.67^{+0.03}_{-0.02}$ & 0 & $80.20\pm4.42$ & $60.49\pm2.95$ & $5731.70\pm1.47$ & $81.78\pm4.09$ & $8.66\pm0.07$ & $7.60^{+0.25}_{-0.24}$ & $2.14^{+0.12}_{-0.12}$ & / & AIS\_201\_1\_49 &  \\
    SMDG0905069+530547 & $0.68^{+0.04}_{-0.03}$ & 0 & $55.43\pm2.38$ & $61.49\pm1.59$ & $4275.57\pm0.79$ & $61.67\pm3.08$ & $8.57\pm0.06$ & $7.82^{+0.27}_{-0.23}$ & $2.50^{+0.26}_{-0.17}$ & $20.25\pm0.07$ & AIS\_80\_1\_6 &  \\
    SMDG0910265+553452 & $0.94^{+0.04}_{-0.05}$ & 0 & $116.16\pm5.69$ & $62.49\pm3.79$ & $13524.49\pm1.90$ & $184.37\pm9.22$ & $9.63\pm0.07$ & $8.70^{+0.38}_{-0.38}$ & $5.53^{+1.08}_{-0.49}$ & / & AIS\_77\_1\_83 &  \\
    SMDG0915558+295527 & $0.43^{+0.03}_{-0.01}$ & 0 & $107.53\pm2.09$ & $63.49\pm1.39$ & $7306.86\pm0.70$ & $102.90\pm5.15$ & $9.16\pm0.06$ & $8.13^{+0.24}_{-0.25}$ & $4.54^{+0.26}_{-0.26}$ & $19.98\pm0.06$ & AIS\_214\_1\_68 &  \\
    \hline
    \end{tabular}

    % 表格注释
    
    \end{threeparttable}
}
 \hspace*{-4cm}% 向左移动 2 cm
\begin{tablenotes}\footnotesize
 \item \textbf{NOTE.} The first 8 rows of this table are shown here. The full table of UDG and LSBs properties is available online in machine-readable format. Column definitions: 
 
 (1) Adopted SMUDGes UDG candidate name; 
 
 (2) Type of galaxies (UDG/LSB; from FASHI/ALFALFA);
 
 (3) Entry number in FASHI catalog or Arecibo General Catalog (AGC);
 
 (4)-(5) J2000 position of optical centroid (deg);
 
 (6)-(7) J2000 position of corresponding \ion{H}{1} detection (deg); 
 
 (8)-(11) Total apparent magnitude in g/r band before galactic extinction (mag) and flag (0=good; 1=extrapolated); 
 
 (12)–(13) Central surface brightness in $g$ band (mag\,arcsec$^{-2}$) and flag;

 (14)–(15) Effective radius in angular units (arcsec) and flag;

 (16)–(17) Minor to major axis ratio and flag;

 (18)–(19) Velocity width of \ion{H}{1} detection (km\,s$^{-1}$);

 (20) Heliocentric systemic velocity (km\,s$^{-1}$);

 (21) Distance from \ion{H}{1} detection (Mpc); 
 
 (22)-(23) Logarithm of \ion{H}{1} mass and stellar mass (log10[M$_\odot$]); 
 
 (24) Physical effective radius (kpc);
 
 (25)-(26) Total apparent magnitude in NUV band (mag) and GALEX Frame;
 
 (27) Redshifts from DESI Data Release 1.
\end{tablenotes}
\end{sidewaystable*}
\end{adjustwidth}

\newpage

\begin{figure*}
\begin{center}
	\includegraphics[width=0.6\textwidth]{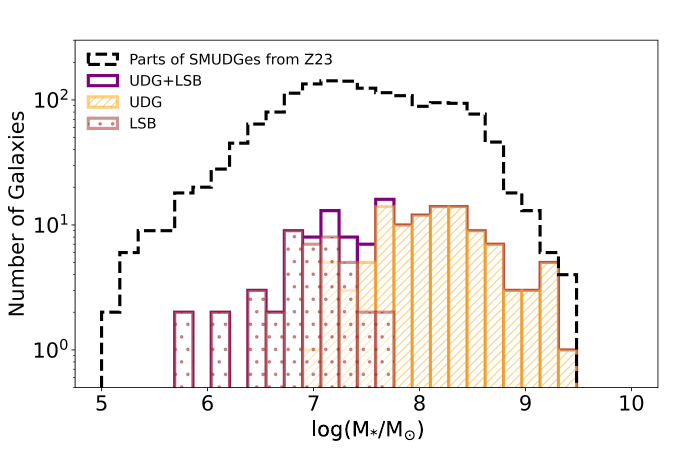}
    \caption{
    Stellar mass distributions for different samples. The dashed outline represents the complete SMUDGes catalog of Z23, restricted to the 1,522 galaxies with available redshift measurements; the solid unfilled bars represent our \ion{H}{1}-detected sample; the dotted region denotes the LSB galaxies; and the hatched region denotes the UDGs.
     }
\label{fig:hist_2}
\end{center}
\end{figure*}

Figure~\ref{fig:hist_2} shows the stellar-mass histogram of our \ion{H}{1}-detected sample, overlaid on the stellar-mass distribution of galaxies in the optically selected SMUDGes catalog of Z23 with available optical redshifts. The stellar-mass distribution of our \ion{H}{1}-detected sample closely resembles that of the redshift available SMUDGes subsample, indicating that our \ion{H}{1}-detected sample is highly representative.

\subsection{Star Formation Rates}
The star formation rate (SFR) serves as an important indicator of a galaxy’s recent star-forming activity and current mass assembly. It can be constrained using a wide range of observational tracers, such as hydrogen recombination lines (e.g., H$\alpha$), radio continuum emission, infrared and ultraviolet luminosities, X-ray emission, or combinations of multi-wavelength measurements. In this work, we derive SFRs from ultraviolet (UV) emission, which is dominated by radiation from massive, short-lived stars and thus traces star formation over relatively recent timescales.
While UV-based SFR estimates are in principle sensitive to dust attenuation, our sample consists of optically faint, blue galaxies with low star formation rates, for which dust effects are expected to be small \citep{da2010new}. We therefore do not apply a dust attenuation correction to the UV luminosities. Under the assumption of negligible dust extinction, the UV luminosity provides an approximately linear proxy for the current SFR.

UV photometry for our sample was obtained from the intensity maps described in Section~2.3. We measured UV fluxes via aperture photometry following a procedure similar to that adopted by \citet{carlsten2022exploration}. Specifically, circular apertures with radii equal to twice the optical effective radius ($2R_{\rm eff}$) were used to capture the bulk of the UV emission. The local background level and associated noise were estimated from an annulus extending from $2R_{\rm eff}$ to $3R_{\rm eff}$. As demonstrated by K24, this aperture-based approach yields results consistent with alternative techniques such as curve-of-growth measurements and random aperture sampling.

The measured fluxes were converted to AB magnitudes using the GALEX zero-points provided by \citet{morrissey2007calibration}. Corrections for Galactic extinction were applied assuming $R_{\rm NUV}=8.2$ and $R_{\rm FUV}=8.24$ \citep{wyder2007uv}. We considered a source to be detected only if its measured magnitude exceeded the corresponding GALEX sensitivity limit; consequently, Table~\ref{table:udg_lsb} lists UV magnitudes exclusively for such detections. Star formation rates were then derived from the UV luminosities using Equations~(5) and (6) of \citet{du2024almost}, which assumed a Salpeter IMF \citep{salpeter1955luminosity}. And they were converted to Chabrier \citep{chabrier2003galactic}, with $\mathrm{SFR}_{\mathrm{Cha}} = \mathrm{SFR}_{\mathrm{Sal}} / 1.41$ \citep{elbaz2007reversal}. Finally, we defined the star formation efficiency (SFE) as the ratio between the SFR and the neutral hydrogen mass, $\mathrm{SFE} = \mathrm{SFR}/M_{\rm HI}$.

\section{Results}
In this section, we investigate scaling relations among the derived galaxy properties and compare them with those reported for the FASHI, ALFALFA $\alpha.100$, and HUDs samples, as well as with the sample presented in K24.

\subsection{\ion{H}{1} and Stellar Mass}
\ion{H}{1} constitutes the principal reservoir for star formation and therefore plays a fundamental role in galaxy evolution. Extensive observational work over the past decades has examined how \ion{H}{1} content relates to a wide range of stellar properties, including galaxy morphology, luminosity, physical size, and star formation activity \citep{disney2008galaxies, garcia2009correlations, west2009h, toribio2011h}. Although the underlying processes linking gas dynamics, star formation, chemical enrichment, and feedback are highly complex, empirical studies have revealed coherent scaling relations connecting the stellar, \ion{H}{1}, and dark matter components of galaxies.

\begin{figure*}
\begin{center}
	\includegraphics[width=0.6\textwidth]{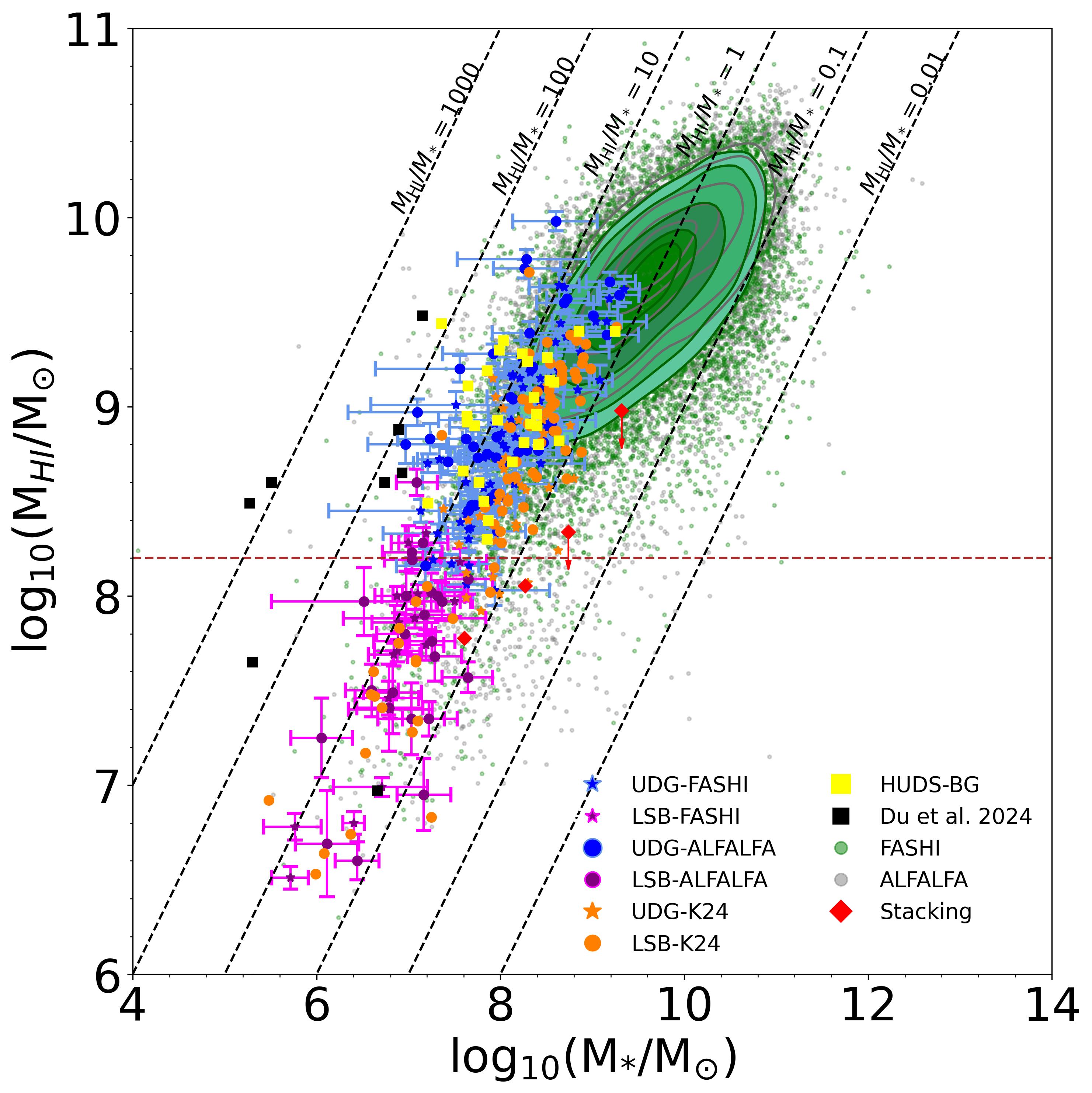}
    \caption{
     The relation between $\log M_{\mathrm{HI}}$--$\log M_{*}$. Orange stars represent UDG-K24, and orange circles represent LSB-K24. Yellow and black squares represent HUDS-BG and \citet{du2024almost}, respectively. Red squares represent the average properties of the bins after stacking, while the red arrows indicate the possible locations of the bins for which no significant \ion{H}{1} signal is detected after stacking. The same symbols used in the following figures represent the same samples as in this figure, unless otherwise noted. The brown dashed line marks $M_{\mathrm{HI}} = 10^{8.2},M_\odot$, which almostly separates UDGs from LSBs.
     Contours in each panel from inside to outside represent 20\%, 40\%,  68\%, 90\%, and 95\% of the sample based on the concentration.
    }
\label{fig:MHI_Ms}
\end{center}
\end{figure*}

Here we selected $\log(\mathrm{M_{HI}})$ and $\log(\mathrm{M_*})$ to investigate the relationship between \ion{H}{1} mass and stellar mass, as shown in Figure~\ref{fig:MHI_Ms}. Our sample (blue and purple points and stars) spans a similar space in this plane as those (orange stars) from K24, and falls toward the low-mass end of the FASHI sample (green points; \citealt{2024SCPMA..6719511Z, 2026arXiv260631539Z}) and the ALFALFA sample (grey points; \citealt{durbala2020alfalfa}). 

We note that our sample is drawn from an optically selected UDG catalogue and may therefore be subject to selection biases against the most diffuse, gas-rich systems. To assess the impact of this effect, we compare our sample with the \ion{H}{1}-selected HUDS-BG sample from \citet{leisman2017almost}, shown as yellow squares in Figure~\ref{fig:MHI_Ms}. The HUDS-BG sample is a subset of the broader HUDS-B galaxies, consisting of 30 objects with GALEX UV imaging within the $\sim40\%$ ALFALFA footprint, and selected based on $\mathrm{r_{eff}} > 1.5~\mathrm{kpc}$, $\mu_{r,\mathrm{eff}} > 24~\mathrm{mag~arcsec^{-2}}$, and $M_r > -17.6~\mathrm{mag}$. As shown in Figure~\ref{fig:MHI_Ms}, our UDGs occupy a similar region in the $M_{\star}$–$M_{\mathrm{HI}}$ plane as the HUDS-BG galaxies, indicating that our sample broadly traces the same \ion{H}{1}-rich UDG population in terms of global gas and stellar mass properties. While selection effects may still be present, they are unlikely to strongly bias the scaling relations explored in this work.

Although FASHI is more sensitive than ALFALFA, the lack of a much larger population of low-$M_{\rm HI}$ systems is not unexpected. Our stellar-mass-dependent comparison is limited to the SMUDGes galaxies with redshift measurements, because both $M_\star$ and $R_{\rm eff}$ require distances. Among the 7070 SMUDGes sources considered here, only 1529 have redshifts, corresponding to a redshift completeness of 21.6\%. This subset is moderately biased toward optically brighter galaxies: 14.7\% of the sources without redshifts have $g>21$, compared to 7.9\% of those with redshifts. Within the redshift-available SMUDGes subset in the \ion{H}{1} survey footprint, the individual \ion{H}{1} detection fraction is only 17.4\% for $10^7<M_\star/M_\odot<10^8$ and 18.7\% for $10^7<M_\star/M_\odot<10^9$. Thus, many optically selected low-stellar-mass diffuse galaxies remain undetected in \ion{H}{1}. This is further supported by the stacking analysis in section 5.2, which shows that  these \ion{H}{1}-undetected systems are not necessarily gas-free, but have average $M_{\rm HI}/M_\star$ values more than 0.5 dex lower than the FASHI- and ALFALFA-detected gas-rich UDGs. Since this offset corresponds to a factor of $\sim3$, the factor of $\sim2$--$3$ sensitivity improvement of FASHI over ALFALFA is not sufficient to reveal a substantially larger population of such relatively gas-poor UDGs, especially given the dependence of the \ion{H}{1} detection limit on distance and linewidth.

\begin{figure*}
\begin{center}
	\includegraphics[width=0.8\textwidth]{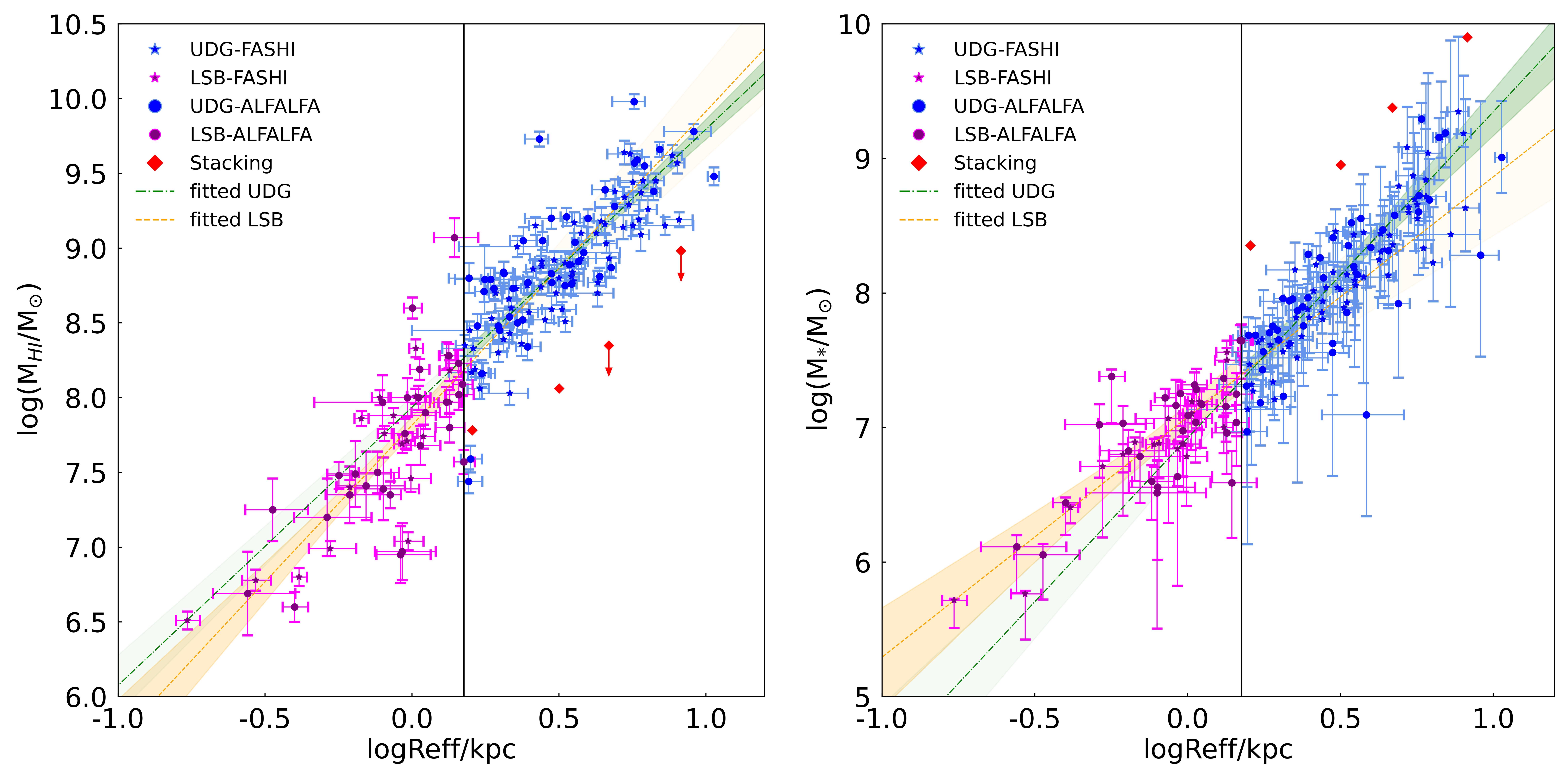}
    \caption{
     Left panel: the relation between $\log M_{\mathrm{HI}}$--$\log R_{eff}$. Right panel: the relation between $\log M_{\mathrm{*}}$--$\log R_{eff}$. The green dash-dotted and orange dashed lines represent the fits to the UDG and LSB distributions, respectively, in both panels, with the shaded regions showing their uncertainties. The black vertical line marks $x=\log(1.5,\mathrm{kpc})$, separating the UDG and LSB regimes in this work. The shaded bands are displayed with higher transparency in the extrapolated regimes, i.e. the UDG fit at $R_{\mathrm{eff}}<1.5,\mathrm{kpc}$ and the LSB fit at $R_{\mathrm{eff}}>1.5,\mathrm{kpc}$, to indicate that these ranges are less certain.
    }
\label{fig:MHI_Ms_Reff}
\end{center}
\end{figure*}

Furthermore, in the $M_{\mathrm{HI}}$–$M_{*}$ plane, the UDGs in our sample lie predominantly above $M_{\mathrm{HI}} = 10^{8.2}~M_\odot$ line, while the LSBs fall below this threshold, showing a clear separation between the two subsamples. For the comparison sample from K24 (orange), the LSB population consists of two subclasses defined using different criteria: one satisfies the same definition adopted in this work, namely $\mu_{0,g} \geq 24~\mathrm{mag~arcsec^{-2}}$ and $R_{\mathrm{eff}} < 1.5~\mathrm{kpc}$, while the other includes sources with $\mu_{0,g} < 24~\mathrm{mag~arcsec^{-2}}$ and $R_{\mathrm{eff}} > 1.5~\mathrm{kpc}$. We find that LSBs in K24 that follow our definition exhibit a distribution consistent with that of our LSBs and UDGs in the $M_{\mathrm{HI}}$–$M_{*}$ plane, whereas the LSBs that deviate from our definition are primarily responsible for the apparent overlap with the UDG population.
In Figure~\ref{fig:MHI_Ms_Reff}, we present the relationships of $\log M_{\mathrm{HI}}$--$\log R_{\mathrm{eff}}$ and $\log M_{*}$--$\log R_{\mathrm{eff}}$. For sources with $u_{0,g} > 24 \ \mathrm{mag \, arcsec^{-2}}$, both $\log M_{\mathrm{HI}}$ and $\log M_{*}$ show a proportional relationship with $\log R_{\mathrm{eff}}$. We further applied the hierarchical Bayesian linear regression algorithm \texttt{LinMix}\footnote{\url{https://github.com/jmeyers314/linmix}}, using its Python implementation \citep{kelly2007some}, to fit the distributions of the UDG and LSB samples separately. In the fitting process, we assumed symmetric uncertainties and adopted the larger value of the upper and lower error bounds for each data point.

We present the first systematic analysis of the relation between \ion{H}{1} mass and optical effective radius for low-surface-brightness galaxies, including both UDGs and LSBs. We found 
\begin{equation}
\log M_{\mathrm{HI}} = (1.774 \pm 0.123)\log R_{\mathrm{eff}} + (7.989 \pm 0.069)
\end{equation}
for the UDG sample, and
\begin{equation}
\log M_{\mathrm{HI}} = (2.013 \pm 0.246)\log R_{\mathrm{eff}} + (7.828 \pm 0.053)
\end{equation}
for the LSB galaxies. The two slopes are consistent with each other within $1\sigma$, indicating similar \ion{H}{1} mass–optical size relations for the two populations. For comparison, \citet{wang2016new} reported a relation for normal disk galaxies of
\begin{equation}
\log M_{\mathrm{HI}} = (1.976 \pm 0.012)\log D_{\mathrm{HI}} + (6.508 \pm 0.042).
\end{equation}
The slopes of the UDG and LSB relations are consistent with that reported by \citet{wang2016new} at the $\sim2\sigma$ and $\sim1\sigma$ levels, respectively.
Despite the different definitions of $R_{\mathrm{eff}}$ (the half-light radius) and $D_{\mathrm{HI}}$ (the \ion{H}{1} diameter defined at a fixed surface density threshold), the broadly similar slopes may reflect the previously reported connection between optical and \ion{H}{1} sizes (e.g. \citealt{1997A&A...324..877B}; \citealt{2002A&A...390..829S}; \citealt{wang2016new}). However, we caution that the global stellar mass--size relation itself exhibits substantial intrinsic scatter and deviates from a single power law over the full galaxy population (e.g. Trujillo et al. 2020), while the ($M_{\mathrm{HI}}$--$M_{\mathrm{*}}$) relation also contains significant scatter. Therefore, the observed relation between ($M_{\mathrm{HI}}$) and optical effective radius should not be interpreted as an intrinsically tight or universal scaling law.
This interpretation is further supported by the four \ion{H}{1}-stacking measurements, which deviate substantially from the relation defined by the directly \ion{H}{1}-detected galaxies and exhibit much larger scatter in the ($M_{\mathrm{HI}}$--$R_{\mathrm{eff}}$) plane. This suggests that the underlying relation is likely broader than implied by the directly detected sample alone, while the comparatively tight separation seen among the detected galaxies may partly reflect the selection biases inherent to \ion{H}{1}-selected surveys, which preferentially detect gas-rich systems.
Nevertheless, the consistency in slope between our \ion{H}{1}-rich diffuse galaxies and normal disk galaxies suggests that the atomic gas distributions of \ion{H}{1}-rich UDGs and LSBs are not fundamentally distinct from those of ordinary galaxies, but instead broadly follow similar empirical size--mass trends. This is broadly consistent with the results of \citet{2021ApJ...909...19G}, who examined the ($M_{\mathrm{HI}}$--$D_{\mathrm{HI}}$) relation for \ion{H}{1}-bearing UDGs and found that, despite their diffuse stellar populations, these galaxies lie on the same relation as normal gas-rich galaxies, with typical global \ion{H}{1} surface densities.

We further examine the relation between stellar mass and optical effective radius. By fitting the $\log M_{\ast}$–$\log R_{\mathrm{eff}}$ plane, we find $M_{\ast} \propto R_{\mathrm{eff}}^{2.441 \pm 0.143}$ for UDGs, and $M_{\ast} \propto R_{\mathrm{eff}}^{1.841 \pm 0.206}$ for LSBs. Given the relatively large uncertainties, the two slopes are consistent with each other within $\sim2\sigma$, and no statistically significant difference between the two populations can be established.

The fitted relations are broadly consistent with $M_\ast \propto R_{\mathrm{eff}}^{2}$ for both UDGs and LSBs. Under this approximation, the corresponding stellar surface mass density, defined as $\mu_{\ast} \sim M_{\ast}/(\pi R_{\mathrm{eff}}^{2}(b/a)^{2})$, would remain approximately constant within the uncertainties. This indicates that UDGs and LSBs with $\mu_{0,g} \gtrsim 24~\mathrm{mag~arcsec^{-2}}$ share similar stellar surface mass densities, although the relation still exhibits substantial scatter, with any residual variations primarily reflecting scatter (including axial ratio effects) rather than a systematic dependence on galaxy size. It also suggests that LSBs with $\mu > 24~\mathrm{mag~arcsec^{-2}}$ represent a type of special LSBs, whose properties—including the surface density distribution—are more similar to those of UDGs. For comparison, \citet{prole2021stellar} reported a slope of $2.512 \pm 0.405$ for dwarf galaxies, which is consistent with both the UDG and LSB results within the uncertainties.
Combined with the $M_{\mathrm{HI}}$–$R_{\mathrm{eff}}$ relation discussed above, these findings suggest that \ion{H}{1}-rich diffuse galaxies broadly follow optical size--mass trends similar to those observed in normal galaxies. However, we caution that the $M_\ast$–$R_{\mathrm{eff}}$ relation itself exhibits substantial intrinsic scatter, and unresolved single-dish \ion{H}{1} measurements do not allow us to directly determine whether UDGs obey the canonical \ion{H}{1} mass--\ion{H}{1} size relation established from resolved data. Therefore, the observed trends should not be interpreted as evidence for a tight or universal scaling law. Instead, our results suggest that UDGs and LSBs likely represent low surface density extensions of ordinary galaxies, at least within the \ion{H}{1}-rich systems probed by the current sample.
In this work, all galaxies in our sample satisfy a surface brightness selection of $u_0, g > 24\,\mathrm{mag\,arcsec^{-2}}$. As a consequence of this selection, the resulting population exhibits a clear separation in effective radius space, as illustrated in Fig.~\ref{fig:MHI_Ms_Reff}. Given this separation, the fitted scaling relations for the two populations are constrained over partially distinct ranges in $R_{\mathrm{eff}}$. The extrapolated portions of the best-fit relations beyond the regions directly populated by the data are shown solely for visual comparison. Their purpose is to assess the consistency of the functional form of the relations between the two populations, rather than to provide physical predictions outside the observationally constrained parameter space.

\subsection{Baryonic Tully-Fisher Relation}
The baryonic Tully–Fisher relation (BTFr), first introduced by \citet{mcgaugh2000baryonic}, links the total baryonic mass $M_{\mathrm{bar}}$ of a galaxy to its characteristic rotation velocity through a power-law scaling of the form $M_{\mathrm{bar}} \propto v^{4}$. Subsequent observational studies have demonstrated that this relation holds over a broad range of galaxy populations (\citealt{ponomareva2021mightee}; \citealt{mcquinn2022turndown}). 

We adopted the method of \citet{du2024almost} to represent the rotation velocity as $V = W_{\mathrm{int}}/2$, where $W_{\mathrm{int}}$ is the intrinsic linewidth of the \ion{H}{1} line profile. 
We adopted $W_{20}$ as the estimator of the intrinsic linewidth, since previous studies have shown that $W_{50}$ is more susceptible to noise and to the blending of multiple velocity components in the \ion{H}{1} line profile, whereas $W_{20}$ is less sensitive to these effects and thus provides a more reliable estimate of the intrinsic linewidth (\citealt{lelli2016sparc}; \citealt{ponomareva2017multiwavelength}). This behavior has also been demonstrated for dwarf galaxies using resolved \ion{H}{1} kinematics from the LITTLE THINGS survey \citep{hu2023global}.
To perform inclination correction, we initially adopted the thin-disk approximation and applied the relations $\cos i = \mathrm{b/a}$ and $W_{\mathrm{int}} = W_{\mathrm{obs}} / \sin i$, where $\mathrm{b/a}$ is the axis ratio from the complete SMUDGes catalog.
We further tested the effect of finite disk thickness by adopting an intrinsic axial ratio of $q_0 = 0.3$, motivated by previous studies suggesting that dwarf and low-surface-brightness galaxies can possess substantially thick disks \citep{2013MNRAS.436L.104R}. In this case, the inclination was calculated using
\begin{equation}
\cos^2 i = \frac{(b/a)^2 - q_0^2}{1 - q_0^2}.
\end{equation}
We find that adopting $q_0 = 0.4$ changes the inclination-corrected linewidths and inferred BTFR offsets by less than $\sim 2\%$, and therefore does not qualitatively affect our conclusions.
We also note that unresolved rotation, beam-smearing effects, and possible mismatches between optical and \ion{H}{1} inclinations may introduce additional uncertainties in the inferred linewidths and inclination corrections, particularly for low-surface-brightness dwarf systems.
Similar to \citet{du2024almost}, we defined the baryonic mass as $M_{\mathrm{bar}} = M_{*} + \eta M_{\mathrm{HI}}$ with the factor $\eta = 1.33$. The estimation of the baryonic mass uncertainty followed Equation~(7) from \citet{du2024almost}. We note that, during this process, we assumed symmetric uncertainties for both the \ion{H}{1} mass and the $g$-band luminosity, and selected the larger value from the lower and upper bounds on these derived quantities.

\begin{figure*}
\begin{center}
	\includegraphics[width=0.6\textwidth]{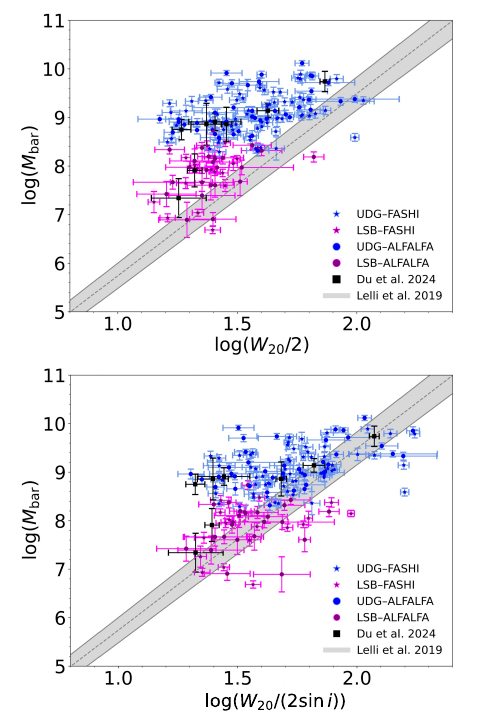}
    \caption{
    Baryonic Tully--Fisher relation of our galaxies and sources from \citet{du2024almost}, using $W_{20}$ as the velocity estimator. 
    Double-peaked galaxies are shown with stars for the FASHI sample and circles for the ALFALFA sample, with blue markers for UDGs and purple markers for LSBs. 
    Single-peaked galaxies are shown with cyan stars for LSBs and cyan circles for UDGs. 
    Sources from \citet{du2024almost} are shown as black squares. 
    The upper panel shows the locations of the sources before optical inclination corrections, while the lower panel shows the locations after the corrections. 
    The BTFr of bright galaxies from \citet{lelli2019baryonic} is shown as a black dashed line, with the corresponding scatter indicated by the gray shaded region.
     }
\label{fig:Mbar_w20}
\end{center}
\end{figure*}

Figure~\ref{fig:Mbar_w20} shows the locations of our sources and sources from \citet{du2024almost} in the BTFr diagram. The upper panel displays the sources without optical inclination correction, while the lower panel shows the results after the optical inclination correction. We also compared our sources with the sample of 121 late-type disk galaxies from SPARC \citep{lelli2016sparc}. Before inclination correction, most of our sources deviate from the empirical BTFr toward the heavier baryonic mass end. After inclination correction, about 30\% of the sources align with the BTFr comparatively, while about half still remain offset to the heavier baryonic mass end, and fewer than 20\% deviate toward the lower baryonic mass end.
We further note that ultra-diffuse galaxies with resolved \ion{H}{1} kinematics also show systematic deviations from the canonical BTFR in the same direction, i.e. toward higher baryonic masses at a given velocity. In particular, \citet{2019ApJ...883L..33M} and \citet{2025A&A...703A.295S} find that UDGs with spatially resolved rotation curves lie above the BTFR when circular velocities are used. This suggests that the offset observed in our sample, based on unresolved \ion{H}{1} linewidths, is qualitatively consistent with results from resolved kinematic studies.
Agreement with the BTFR indicates that these galaxies follow the empirical scaling relation between baryonic mass and characteristic velocity observed in rotationally supported systems. This is consistent with a coupling between the baryonic content and the underlying gravitational potential.
We stress that the BTFR alone does not provide direct constraints on detailed evolutionary pathways such as merger history or the angular momentum evolution of individual galaxies. Deviations toward the lower baryonic mass end may arise from gas depletion processes, such as ram-pressure stripping \citep{mcgaugh2010local}. In contrast, offsets toward the higher baryonic mass end may have multiple physical origins. These include environmental effects such as tidal interactions, but also internal processes such as inefficient feedback that fails to expel gas from the disk, as suggested in resolved studies of ultra-diffuse galaxies \citep{2020MNRAS.495.3636M, 2021ApJ...909...20S}. 
High-angular-momentum halos may contribute to the formation of spatially extended, low-surface-density disks \citep{2016MNRAS.459L..51A}. Independently, low-surface-brightness galaxies are often observed to exhibit slowly rising rotation curves \citep{1996MNRAS.283...18D}. However, a comparison with the small number of UDGs currently having suitable spatially resolved \ion{H}{1} kinematics \citep{2020MNRAS.495.3636M,2021ApJ...909...20S} provides no evidence that inclination-corrected $W_{20}$-based velocities systematically underestimate their resolved outer rotation velocities. Given the limited number of resolved systems and the possible sensitivity of $W_{20}$ to low-level profile wings and non-rotational broadening, a larger resolved \ion{H}{1} sample is required to determine how reliably global linewidths trace the outer parts of UDG rotation curves. The magnitude of the inferred BTFR offset should therefore be interpreted with caution.

We note that the scatter of the BTFr is known to increase for dwarfs due to the larger uncertainties encountered in measuring their line width and magnitude (\citealt{begum2008baryonic}, \citealt{mcgaugh2012baryonic}. Moreover, owing to the low surface brightness and irregular morphologies of our sources, the uncertainties in inclination are substantial. Additionally, inclinations derived from optical bands may not accurately represent those of the \ion{H}{1} gas \citep{trachternach2009baryonic,2024A&A...689A.344M}. Nevertheless, these inclination uncertainties are unlikely to fully invalidate the overall trend of the sources toward the heavier baryonic mass.

We also note that we do not apply an asymmetric drift correction to the \ion{H}{1} linewidths. We performed a simple test to estimate the potential impact of asymmetric drift by applying a first-order correction of the form $V_{\rm corr} = \sqrt{V^2 + \sigma^2}$, assuming a characteristic \ion{H}{1} velocity dispersion of $\sigma \sim 10\,{\rm km\,s^{-1}}$ typical of dwarf and low-surface-brightness galaxies. We find that this correction leads to only modest changes in the inferred velocities and does not qualitatively affect the position of our sources relative to the BTFR. This is consistent with previous studies \citep{2017MNRAS.466.4159I, 2025A&A...699A.311M}, which also find that asymmetric drift corrections are insufficient to fully reconcile dwarf galaxies with the BTFR.

\subsection{Star Formation Rate and Star Formation Efficiency}
Galaxies selected via their neutral hydrogen content, including those analyzed in this work, are predominantly blue and are commonly associated with ongoing or recently sustained star formation activity \citep{huang2012arecibo}. Blue optical colors are widely interpreted as signatures of active star formation. For such star-forming systems, observations have established a stellar-mass-dependent relation in which the star formation rate increases with stellar mass up to a characteristic scale and declines at higher masses \citep{noeske2007star,daddi2007multiwavelength}. This relation is commonly referred to as the star-forming main sequence (MS) and has been shown to hold over a broad range of cosmic epochs.

In contrast, LSBs are found to depart systematically from the MS. A large body of observational work demonstrates that LSB systems tend to lie below the MS defined by higher-surface-brightness, more massive galaxies, exhibiting suppressed star formation rates at fixed stellar mass \citep{van1993star,boissier2008galex,wyder2009star,meurer2009evidence,schombert2011stellar,schombert2013stellar,schombert2014stellar,schombert2015stellar}. This systematic offset is generally attributed to inefficient star formation in diffuse, low-density environments, and is often interpreted as a consequence of reduced star formation efficiency (SFE) in LSB galaxies \citep{hunter2010galex,geha2012stellar,huang2012gas,lei2018halpha}.

To place our results in context, we constructed two comparison samples following the approach of \citet{huang2012arecibo}: the $\alpha.100$–GALEX–SDSS and FASHI–GALEX–SDSS datasets. The $\alpha.100$–GALEX–SDSS sample was obtained by cross-matching the $\alpha.100$–SDSS catalog \citep{durbala2020alfalfa} with the GALEX–SDSS UV–optical catalog of \citet{bianchi2017revised}, which combines GALEX AIS GUVcat data with SDSS DR14. For the FASHI-based comparison sample, we first matched the FASHI catalog to SDSS following the criteria of \citet{2024SCPMA..6719511Z, 2026arXiv260631539Z}, and then cross-matched the resulting catalog with GALEX–SDSS. The final comparison samples comprise 18,187 galaxies in the $\alpha.100$–GALEX–SDSS sample and 14,395 galaxies in the FASHI–GALEX–SDSS sample, respectively.

\begin{figure*}
\begin{center}
	\includegraphics[width=0.6\textwidth]{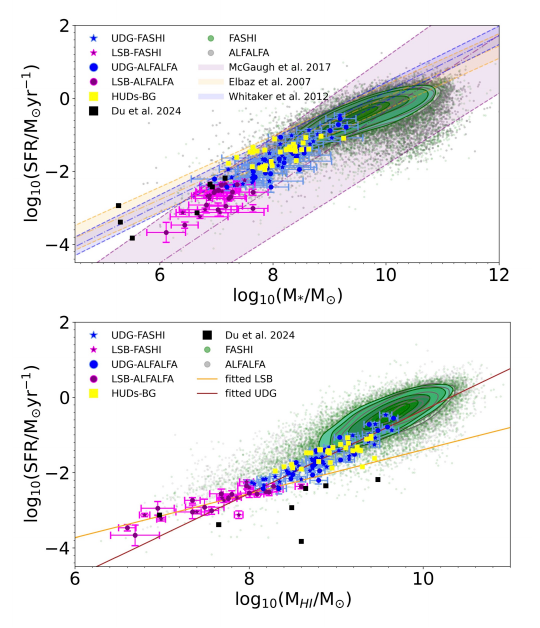}
    \caption{
    Upper: the relation between the star formation rate and $M_{*}$ for our sources, together with comparisons to the FASHI sample, the $\alpha.100$ sample, HUDs-BG, and MSs. The FASHI--GALEX--SDSS and $\alpha.100$--GALEX--SDSS samples are shown as green and gray contours and points, respectively, with contour levels in each panel identical to those in Figure~\ref{fig:MHI_Ms}. Three empirical MSs and their scatters, from \citealt{elbaz2007reversal}, \citealt{whitaker2012star}, \citealt{mcgaugh2017star}, are indicated by orange, blue, and purple dashed lines with shaded regions, respectively. Lower: the relation between SFR and $M_{\rm HI}$ for our sources, compared with the FASHI sample, the $\alpha.100$ sample, and HUDs-BG. Symbols are the same as in the upper panel. The orange and brown solid lines correspond to the best-fitting linear relations derived for the LSB and UDG samples, respectively.
     }
\label{fig:sfr_Ms_MHI}
\end{center}
\end{figure*}

The upper panel of Figure~\ref{fig:sfr_Ms_MHI} shows the $\log \mathrm{SFR}$ versus stellar mass relation for our sample galaxies, and comparing with the HUDS-BG sample from \citet{leisman2017almost}, as well as with the FASHI--GALEX--SDSS and $\alpha.100$--GALEX--SDSS samples. We used the same method described in Section~3.2 to estimate the SFRs of the FASHI--GALEX--SDSS and $\alpha.100$--GALEX--SDSS samples, in order to ensure the validity of the comparison. For further context, our sample was compared with two reference star-forming main sequences (MSs) at $z=0$. The first MS is defined by blue, star-forming galaxies with no contribution from active galactic nuclei \citep{elbaz2007reversal}, while the second is derived from a large sample of 22,816 star-forming systems—spanning both blue and red populations—drawn from the NOAO Extremely Wide-Field Infrared Imager Medium-Band Survey \citep{whitaker2012star}. In addition, we include the MS relation for late-type low-surface-brightness galaxies presented by \citet{mcgaugh2017star}, which is based on 56 systems with stellar masses between $5\times10^{6}$ and $7\times10^{9},M_\odot$. The UDGs in our sample are consistent with the relationship of star-forming galaxies and exhibit a distribution similar to that of the HUDS-BG sample. However, the LSBs in our sample show lower stellar masses as well as SFR, which align with the trends observed in late-type LSBGs. We plotted our sources on the $\log M_{\mathrm{HI}}$ and $\log \mathrm{SFR}$ relation diagram in the lower panel of Figure~7. Most of our sources exhibit lower SFRs compared to the FASHI--GALEX--SDSS sample and $\alpha.100$--GALEX--SDSS sample. Meanwhile, the positions of our UDGs are consistent with that of the HUDS-BG sample on the diagram.

\begin{figure*}
\begin{center}
	\includegraphics[width=0.6\textwidth]{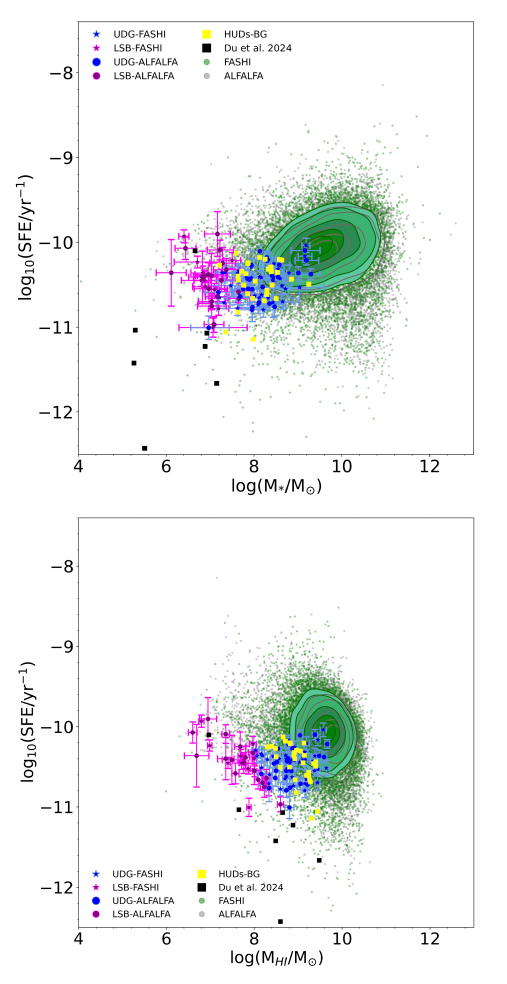}
    \caption{
    Upper: the relation between $\log(\mathrm{SFE})$ and $\log(M_{*})$. Lower: the relation between $\log(\mathrm{SFE})$ and $\log(M_{HI})$.
     }
\label{fig:SFE_Ms_MHI}
\end{center}
\end{figure*}

Compared with the sources from the FASHI--GALEX--SDSS sample and the $\alpha.100$--GALEX--SDSS sample of similar $M_{\mathrm{HI}}$, our galaxies show lower SFRs, pointing to inefficient conversion of \ion{H}{1} into stars. This may reflect a reduced atomic-to-molecular transition rate \citep{cao2017molecular}, or an extended gas distribution where the surface density remains below the star formation threshold \citep{kennicutt1989star, martin2001star}. Additional factors, including gas stability and ISM properties, could further suppress the star formation efficiency (SFE $=$ SFR/$M_{\mathrm{HI}}$). Thus, we plotted $\log$(SFE) versus $M_{*}$ and  $\log$(SFE) versus $M_{HI}$ in Figure~\ref{fig:SFE_Ms_MHI}. The SFE of our sources is clustered around $\sim10^{-10.5}\,\mathrm{yr}^{-1}$, corresponding to a current gas consumption time (Roberts time, $t_{R} = M_{\mathrm{HI}} / \mathrm{SFR}$) of about 32~Gyr, which is comparable to that of HUDS \citep{leisman2017almost}. In contrast, the optically selected GALEX Arecibo SDSS Survey (GASS) sample, a stellar-mass–selected survey of nearby galaxies, has an average $t_{R}$ of 3~Gyr \citep{schiminovich2010galex}, while the averages for FASHI and ALFALFA are $\sim8.9$~Gyr. This difference cannot be attributed solely to a selection effect, since $t_{R}$ is found to be nearly independent of $M_{*}$ or $M_{\mathrm{HI}}$ \citep{huang2012gas}. We also note that several LSB galaxies in our sample exhibit seemingly higher SFEs than UDGs at the same stellar mass. This is primarily a ratio effect arising from their extremely low $M_{\mathrm{HI}}$ and SFRs, rather than indicating genuinely enhanced star formation activity.
These cases arise because they occupy the lower-left region of the $\log$(SFR)–$\log(M_{\mathrm{HI}})$ diagram, with extremely low $M_{\mathrm{HI}}$ and SFRs.

In the lower panel of Figure~\ref{fig:sfr_Ms_MHI}, we similarly applied \texttt{LinMix} to perform separate best-fitting linear regressions for the LSB and UDG samples. We found that the two populations follow distinct relations with significantly different slopes, namely $\mathrm{SFR} \propto M_{\mathrm{HI}}^{0.588 \pm 0.073}$ for LSBs and $\mathrm{SFR} \propto M_{\mathrm{HI}}^{1.115 \pm 0.060}$ for UDGs.
We note that UDGs in our sample are systematically more \ion{H}{1}-rich than LSB galaxies and therefore occupy a higher $M_{\mathrm{HI}}$ regime. 
This difference is intrinsic to the two populations, and restricting the fits to the same $M_{\mathrm{HI}}$ range would exclude a large fraction of UDGs and obscure the physical distinction between them.
Consequently, the implied star formation efficiency scales as $\mathrm{SFE} \propto M_{\mathrm{HI}}^{-0.412 \pm 0.072}$ for LSBs, while for UDGs the inferred $\mathrm{SFE}$--$M_{\mathrm{HI}}$ relation is consistent with being flat within the uncertainties. This naturally explains the behavior seen in the lower panel of Figure~\ref{fig:SFE_Ms_MHI}, where LSB galaxies exhibit a clear anti-correlation between $\log \mathrm{SFE}$ and $\log M_{\mathrm{HI}}$, whereas no obvious scaling relation is observed for UDGs.

The UDGs in our sample have relatively low $M_{*}$ but contain large reservoirs of \ion{H}{1}. In the $M_{\mathrm{HI}}$–$\log(\mathrm{SFR})$ plane, their SFRs lie slightly above the MS of LSBGs, yet remain lower than those of the FASHI and $\alpha.100$ samples. These properties indicate that UDGs are currently inefficient at forming stars despite their abundant atomic gas, potentially reflecting delayed or episodic star formation regulated by multiple physical processes. Alternatively, if UDGs indeed undergo bursty star formation histories as proposed by \citet{di2017nihao}, we may be observing them during a phase of significant gas accumulation, prior to a substantial increase in their star formation rate. We will discuss the formation and evolution of UDGs in detail in Section~5.3.

\subsection{\ion{H}{1} stacking results}
\begin{figure*}
\begin{center}
	\includegraphics[width=0.8\textwidth]{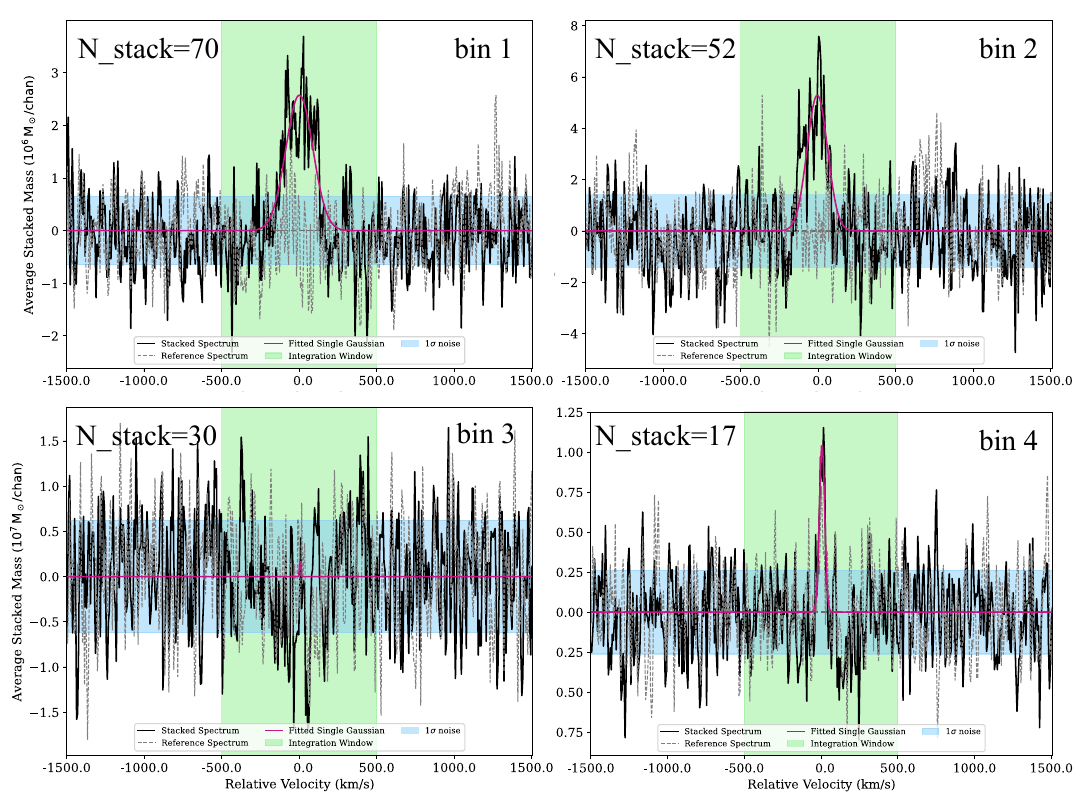}
    \caption{
    Stacked \ion{H}{1} spectra for four stellar mass bins. The sample was divided into Bin 1 ($6.0 < \log(M_\ast/M_\odot) < 8.0$), Bin 2 ($8.0 < \log(M_\ast/M_\odot) < 8.5$), Bin 3 ($8.5 < \log(M_\ast/M_\odot) < 9.0$), and Bin 4 ($9.0 < \log(M_\ast/M_\odot) < 9.6$). 
    The number in the upper-left corner of each panel ($N_{\rm stack}$) indicates the number of galaxies included in that bin. 
    The stacked \ion{H}{1} spectra do not show double-peaked profile, but single-peaked. So we fit the \ion{H}{1} mass by gaussian fitting. The spectrum channel width is 6.4~km~s$^{-1}$.
}

\label{fig:stacking}
\end{center}
\end{figure*}

The average \ion{H}{1} spectra from stacking for the galaxies in the four bins are shown in Figure~\ref{fig:stacking}. Each panel indicates the number of galaxies included in the stack ($N_{\rm stack}$) in the upper-left corner. Clear \ion{H}{1} signals are detected in the stacked spectra of bin~1 and bin~2, whereas bin~3 and bin~4, which have smaller sample sizes, do not show significant \ion{H}{1} detection. Although a sharp peak appears in the average \ion{H}{1} spectrum of bin~4, its extremely narrow velocity width suggests that it is likely dominated by a few strong emitters rather than representing the typical \ion{H}{1} content of this bin.  

We note that the signal-to-noise ratio (S/N) of stacked spectra approximately scales as $\sqrt{N_{\rm stack}}$, and therefore the lack of significant detection in bin~3 and bin~4 could be partially due to their smaller sample sizes, in addition to an intrinsically low \ion{H}{1} content. For the bins without significant \ion{H}{1} detections, we estimated upper limits on the \ion{H}{1} mass based on the $3\sigma$ noise level in the stacked spectra. Table~\ref{table:stacking_properties} summarizes the average properties of the four bins, including $N_{\rm stack}$ and the derived \ion{H}{1} masses or upper limits.  

\begin{table*}
    \centering
    \caption{The average properties of the four stacking bins} 
     \label{table:stacking_properties} 
    \begin{tabular}{c|cccc}
        \toprule % 顶部粗横线
         & N\_stack & $\log M_\mathrm{HI}$ & $\log M_\mathrm{*}$ & $R_\mathrm{eff}$ \\
         & & log[M$_\odot$] & log[M$_\odot$] & kpc \\
        (1) & (2) & (3) & (4) & (5) \\
        \hline % 中间横线
        bin1 ($6.0 < \log(M_\ast/M_\odot) < 8.0$) & 70 & 7.78 & 7.61 & 1.61 \\
        bin2 ($8.0 < \log(M_\ast/M_\odot) < 8.5$) & 52 & 8.05 & 8.27 & 3.18 \\
        bin3 ($8.5 < \log(M_\ast/M_\odot) < 9.0$) & 30 & $<$8.34 & 8.74 & 4.68 \\
        bin4 ($9.0 < \log(M_\ast/M_\odot) < 9.6$) & 17 & $<$8.98 & 9.32 & 8.21 \\
        \hline % 底部粗横线
    \end{tabular}
\end{table*}

\begin{figure*}
\begin{center}
	\includegraphics[width=0.6\textwidth]{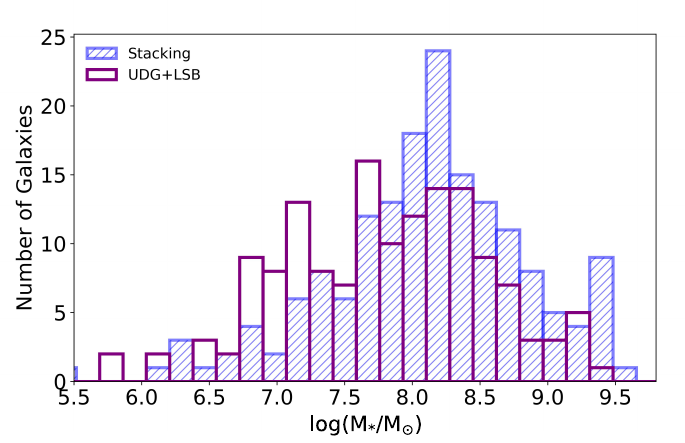}
    \caption{
    Stellar mass distributions for different samples. The unfilled region represents our \ion{H}{1} detected sample, while the hatched region represents the galaxies used in our stacking analysis.
     }
\label{fig:hist_1}
\end{center}
\end{figure*}

Figure~\ref{fig:hist_1} shows the stellar mass histograms of both the \ion{H}{1} detected and nondetected samples used for stacking. About 83.2\% of the \ion{H}{1}-detected sample lies within $\log M_\ast < 8.5$, while 44.9\% of the \ion{H}{1}-nondetected sample have $\log M_\ast > 8.5$, consistent with the well-known trend that the \ion{H}{1} fraction decreases with increasing stellar mass.  

Figures~\ref{fig:MHI_Ms} and \ref{fig:MHI_Ms_Reff} show the locations of the four bins in the corresponding scaling relations. Bins with significant stacked \ion{H}{1} detections (bin~1 and bin~2) are broadly consistent with galaxies that have individual \ion{H}{1} detections. Bins without significant detections (bin~3 and bin~4) lie systematically below the distribution of HI-detected galaxies at fixed stellar mass or $R_{\rm eff}$, supporting the conclusion that UDG candidates without \ion{H}{1} detections are generally HI-poor. We caution, however, that part of the non-detection in these bins may be influenced by their smaller $N_{\rm stack}$, which reduces the S/N of the stacked spectra.

\section{Discussion}
In the following sections, we first discuss the selection effects and completeness of our sample (Section~5.1), then examine the gas fraction scaling at the low-$M_{*}$ end (Section~5.2), and finally evaluate our results in the context of the bursty star formation mechanism (Section~5.3).

\subsection{Selection Effects and Completeness}
Our sample was constructed by cross-matching \ion{H}{1} blind surveys (FASHI and ALFALFA) with the SMUDGes catalog, and was therefore not an unbiased representation of the overall UDG/LSB population. The SMUDGes catalog was derived from Legacy Surveys DR9 imaging and applies selection thresholds of $\mu_{0,g} \geq 24~\mathrm{mag~arcsec^{-2}}$ and $r_{e} \geq 5.3^{\prime\prime}$ (Z23). Because detection depends on central angular half-light radius limits, galaxies with small $r_{e}$ are systematically underrepresented, while sources projected near bright objects or in regions of complex backgrounds are also prone to being missed. At the same time, the catalog becomes increasingly incomplete for extremely diffuse systems with surface brightnesses close to the survey detection limit, even within the UDG/LSB selection regime. Artificial source simulations in Z23 further indicated that this catalog becomes highly incomplete for $r_{e} \gtrsim 20^{\prime\prime}$, and the mean overall completeness for galaxies like those already in the catalog is close to 50\%. 

For the \ion{H}{1} data, both FASHI and ALFALFA naturally favor gas-rich systems. 
In addition to the total \ion{H}{1} flux sensitivity, the detectability of galaxies may also depend on the \ion{H}{1} column density distribution. Galaxies with very extended and diffuse \ion{H}{1} disks could remain undetected if their peak column densities fall below the effective sensitivity limits of ALFALFA or FASHI, even when their total \ion{H}{1} masses are not exceptionally low. This effect may preferentially impact extremely diffuse UDGs. However, our sample still contains numerous galaxies with $M_{\mathrm{HI}} < 10^{8.2}\,M_\odot$, most of which are classified as LSB galaxies because their optical effective radii remain below the UDG threshold. Therefore, low-\ion{H}{1}-mass galaxies themselves are not absent from the survey. Moreover, similarly diffuse low-column-density \ion{H}{1} structures could also affect non-UDG LSB galaxies. Consequently, although selection effects may contribute to the apparent separation between UDGs and non-UDGs in $M_{\mathrm{HI}}$, they are unlikely to fully account for the observed trend.
Moreover, in our cross-matching procedure, galaxies in dense group environments were excluded in order to ensure unique \ion{H}{1}--optical counterpart associations. As a result, our final sample is strongly biased toward gas-rich systems in low-density environments, whereas cluster or group galaxies that are \ion{H}{1}-poor are typically absent. 

The \ion{H}{1}-stacking  sample covers some of the gas-poor UDGs. This sample, combined with  the HI-detected sample, is well representative of the 1522 SMUDG galaxies with optical redshift available, as can be seen in the histogram in  Figure~\ref{fig:hist_2}. 

\subsection{Gas Fraction Scaling at the Low-$M_*$ End}
In the study of diffuse galaxies, one of the most informative scaling relations is the variation of the \ion{H}{1} fraction ($M_{\rm HI}/M_*$) with stellar mass. This relation directly reflects the efficiency with which galaxies convert their gas reservoirs into stars, while simultaneously manifesting the intricate interplay between gas accretion, star formation, and feedback processes. \citet{huang2012arecibo} observed a clear trend in ALFALFA galaxies in which $\log(M_{\rm HI}/M_*)$ decreases with increasing $M_*$ over the range $8.0 < \log M_* < 11.0$. At the high-stellar mass end, the xGASS results (GALEX Arecibo SDSS Survey; \citealt{catinella2010galex}) as well as those of \citet{cortese2011effect}, which focused on trends among galaxies in different environments, similarly revealed a systematic decline of $\log(M_{\rm HI}/M_*)$ with stellar mass, indicating that more massive systems exhibit higher star formation efficiencies and lower gas contents. At the low-stellar mass end ($\log M_* < 9$), however, only the ALFALFA dwarfs trace the flattening off of $\log(M_{\rm HI}/M_*)$ \citep{huang2012arecibo}.

\begin{figure*}
\begin{center}
	\includegraphics[width=0.6\textwidth]{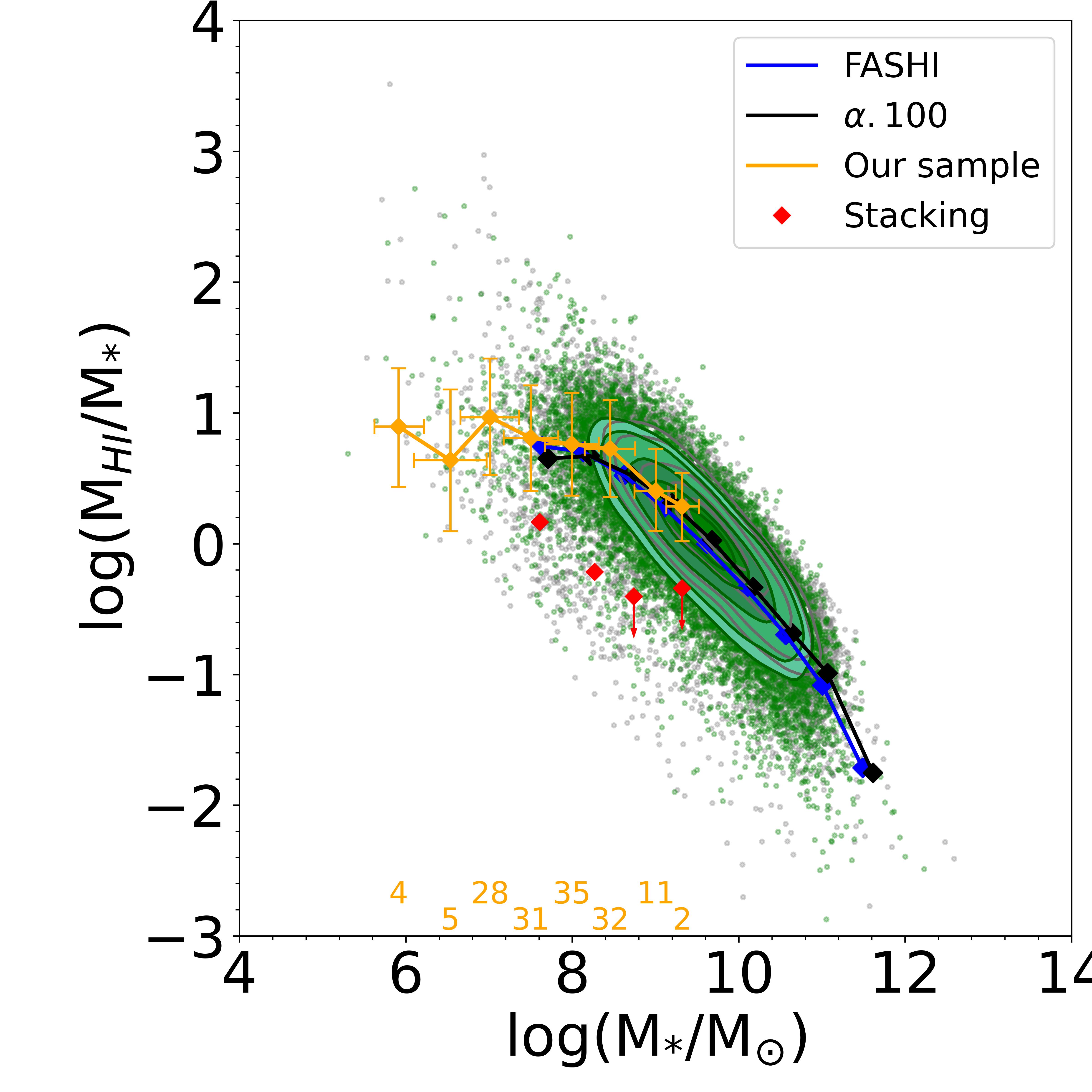}
    \caption{
    The ratio between $\log(M_{\rm HI}/M_{*})$ and $M_*$ is shown. The green and gray contours, scatter points, red squares, and arrows follow the definitions given in the Figure~\ref{fig:MHI_Ms}. Diamonds and solid lines indicate the mean values $\langle \log y \rangle$ in bins of $\log x$, where blue, black, and orange correspond to the FASHI, ALFALFA, and our sample, respectively. The number of galaxies in each $\log x$ bin for our sample is listed at the bottom of the figure.
     }
\label{fig:low_Ms}
\end{center}
\end{figure*}

In Figure~\ref{fig:low_Ms}, the diamonds and solid lines represent the average values $\langle \log(M_{\rm HI}/M_*) \rangle$ and $\langle \log M_* \rangle$ in bins of $\log M_*$ with a bin size of 0.5~dex, where yellow, blue, and black denote our LSB/UDG sample, the FASHI, and ALFALFA sample, respectively. Red square represent the stacking bins. To highlight the overall trend more clearly, we do not plot the individual data points of our sample. Both FASHI and ALFALFA show the same pattern: $\log(M_{\rm HI}/M_*)$ decreases with stellar mass at the high-stellar-mass end, but gradually flattens below $M_* \sim 10^9\,M_\odot$. Our sample, which focuses on LSB/UDG with even lower $M_{\rm HI}$ and $M_*$, extends this flattening trend further toward the low-stellar-mass regime. This may have been influenced by the selection of \ion{H}{1}, and may miss part of the gas-poor galaxy population. Nevertheless, it still demonstrates that LSB/UDGs maintain relatively stable \ion{H}{1} gas fractions at the low stellar mass end.

This behavior appears to be influenced by two main factors: (1) the shallow potential wells of dwarf galaxies may make them more susceptible to stellar feedback, which seems to limit the efficiency of gas-to-star conversion and may help maintain a substantial \ion{H}{1} reservoir \citep{dekel1986origin, somerville2015physical}; and (2) such galaxies typically possess low metallicities and low-pressure environments, which appear to reduce the formation of molecular gas, leaving much of the cold gas in the atomic phase \citep{krumholz2013star, tacconi2020evolution}. As a result, \ion{H}{1} mass appears to scale roughly with stellar mass, yielding an approximately constant $\log(M_{\rm HI}/M_*)$ ratio at the low-stellar-mass end. This conclusion is generally consistent with results from \ion{H}{1} surveys \citep{huang2012arecibo, catinella2018xgass}. However, we note that, due to the limited size of our sample, there is still significant uncertainty in the trend at the very low stellar-mass end.

\subsection{Formation and Evolution of UDGs}
The bursty star formation mechanism proposed by \citet{di2017nihao} using the NIHAO simulations provides one possible framework for understanding the formation of UDGs. In this scenario, repeated episodes of star formation feedback can redistribute baryons and dark matter, leading to the expansion of the stellar component and the formation of extended, low-surface-brightness galaxies. Di Cintio et al. also predicted that UDGs with larger optical sizes tend to have higher gas fractions.

We emphasize, however, that a positive connection between optical size and gas fraction is not unique to the bursty-feedback scenario. Similar trends may also arise in high-angular-momentum systems, where extended disks can retain larger gas reservoirs \citep{2021A&A...651L..15M, 2022MNRAS.516.4043H}, and in other simulation frameworks in which larger UDGs are also predicted to have higher gas fractions \citep{wright2021formation}. Therefore, the observed combination of large $R_{\rm eff}$ and high gas fractions should be interpreted as a general feature of gas-rich extended systems, rather than as unique evidence for a specific formation model.

In our sample, UDGs are characterized by relatively large $R_{\rm eff}$ and modest star formation rates, which together imply low star formation rate surface densities. Similar characteristics have been observed in other UDG samples by \citet{mancera2020robust}, who attributed this trend to weak and inefficient internal feedback, such as supernova-driven feedback.

Furthermore, our sample shows narrow velocity widths (with an average $W_{20}$ of 80.0~km,s$^{-1}$, compared to 194.5~km,s$^{-1}$ and 215.0~km,s$^{-1}$ for the FASHI and ALFALFA samples, respectively) together with large $R_{\rm eff}$.

This combination of large optical sizes and relatively narrow HI velocity widths is qualitatively consistent with systems characterized by relatively high specific angular momentum. In such systems, angular momentum support can contribute to the formation of more extended, lower surface density disks; however, the mapping between angular momentum and observed rotation velocity is not one-to-one and depends on the detailed mass distribution and baryonic response.

In particular, for a given baryonic or halo mass, the halo spin parameter is related to the product of characteristic size and rotation velocity (i.e., $\lambda \propto R V$), and therefore does not uniquely imply that systems with larger sizes must exhibit lower rotation velocities. Both $R$ and $V$ may vary depending on halo structure, baryonic contraction, and formation history. Consequently, high-spin halos do not generically predict low rotation velocities at fixed mass, but may instead produce a range of structural and kinematic configurations.
Therefore, the observed properties are better interpreted as being broadly consistent with a scenario in which a fraction of UDGs reside in relatively high-spin dark matter halos, rather than providing direct or unique evidence for such an origin \citep{amorisco2016ultradiffuse, rong2017universe, 2022ApJ...936..166K}.
However, as noted by \citet{wright2021formation}, the initial spin of a halo does not necessarily correspond to its present-day spin. These model predictions also appear applicable to LSBs at the margins of the UDG definition, implying that multiple formation mechanisms may be at play.

\section{Conclusion}
In this work, we have systematically investigated the \ion{H}{1} properties, stellar content by combining the SMUDGes catalog with FASHI, ALFALFA, and DESI DR1 data. Our main findings are summarized as follows:

\textbf{1. A large statistical sample of \ion{H}{1}-confirmed UDGs:}
We constructed an \ion{H}{1}-detected sample of 160 diffuse galaxies, including 112 UDGs and 48 LSBs, and provided \ion{H}{1}-based redshifts for 76 galaxies for the first time. The distributions of the \ion{H}{1}-detected sample are broadly consistent with those of the parent SMUDGes catalog restricted to galaxies with available redshifts, suggesting that our sample is not strongly biased in terms of the main optical properties considered here. This work therefore establishes one of the largest currently available samples of \ion{H}{1}-confirmed UDGs under the adopted selection criteria ($\mu_{0,g} \geq 24~\mathrm{mag~arcsec^{-2}}$ and $R_{\mathrm{eff}} \geq 1.5~\mathrm{kpc}$).

\textbf{2. Representative \ion{H}{1} stacking constraints from 168 non-detected diffuse galaxies:}
Using DESI DR1 redshifts, we constructed a large and statistically representative sample of 168 diffuse galaxies without individual \ion{H}{1} detections and performed \ion{H}{1} stacking analyses to probe their average gas content below the detection limit. Significant stacking signals are detected in the two lowest stellar-mass bins, while no statistically significant signal is found at higher stellar masses. These results provide robust statistical constraints on the average \ion{H}{1} content of low-mass diffuse galaxies.

\textbf{3. A diffuse LSB population with $\mu_{0,g}\gtrsim24~\mathrm{mag~arcsec^{-2}}$ sharing UDG-like scaling relations:}
Our \ion{H}{1}-detected galaxies extend the low-mass end of the FASHI and ALFALFA populations. Focusing on LSB galaxies with central surface brightness $\mu_{0,g}\gtrsim24~\mathrm{mag~arcsec^{-2}}$, we identify a diffuse LSB subset that closely resembles UDGs in both structural and gas properties. We present a systematic analysis of the relation between \ion{H}{1} mass and optical effective radius for diffuse galaxies, showing that both UDGs and these low surface-density LSBs follow scaling relations consistent with those of normal galaxies, consistent with the absence of strong variation in the average \ion{H}{1} surface density across galaxy populations, as inferred from global scaling relations. Their stellar mass--size relation further indicates nearly constant stellar surface mass densities in this diffuse regime.

\textbf{4. Low star formation efficiency and diverse kinematics:}
Our galaxies show significant scatter relative to the canonical baryonic Tully–Fisher relation, with a tendency toward higher baryonic masses at fixed linewidth-based velocities. This offset is qualitatively consistent with trends seen in resolved UDG kinematic studies, although uncertainties in linewidth measurements and inclination corrections prevent a unique physical interpretation of the deviation.
Both UDGs and LSBs show low star formation efficiencies and long gas depletion times. LSBs display a clear anti-correlation between SFE and \ion{H}{1} mass, whereas UDGs show little dependence, suggesting inefficient and regulated star formation in gas-rich diffuse systems.

\textbf{5. Multiple formation pathways likely contribute to UDGs:} 
The gas fraction at the low-stellar-mass end appears to tend to flatten, indicating that \ion{H}{1} mass scales roughly with stellar mass. Some UDGs in our sample are consistent with bursty star formation, in which intermittent star formation episodes drive cycles of gas inflow and outflow. Their large effective radii, modest star formation rates, and narrow velocity widths suggest weak internal feedback. These results indicate that multiple mechanisms, including bursty star formation and halo dynamics, likely contribute to the observed properties of UDGs, with similar trends extending to LSBs near the UDG definition.

Future observations with deeper \ion{H}{1} and optical data, as well as high-resolution studies of \ion{H}{1} and molecular gas, will be essential to probe lower-mass and gas-poor UDGs, investigate internal dynamics, and assess the roles of feedback, angular momentum, and environmental effects in shaping UDG formation and evolution.

%\begin{acknowledgments}
\section*{Acknowledgments}
We thank the anonymous referee for constructive comments and suggestions, which  significantly improved the clarity and quality of the paper.
This work is supported by the National SKA Program of China (2025SKA0150101) and by the Guizhou Provincial Science and Technology Projects (QKHFQ[2023]003, QKHPTRCZDSYS[2023]003, QKHFQ[2024]001-1). FAST is a Chinese national mega-science facility operated by the National Astronomical Observatories of the Chinese Academy of Sciences (NAOC).

The Legacy Surveys consist of three individual and complementary projects: the Dark Energy Camera Legacy Survey (DECaLS; Proposal ID \#2014B-0404; PIs: David Schlegel and Arjun Dey), the Beijing--Arizona Sky Survey (BASS; NOIRLab Proposal ID \#2015A-0801; PIs: Zhou Xu and Xiaohui Fan), and the Mayall $z$-band Legacy Survey (MzLS; Proposal ID \#2016A-0453; PI: Arjun Dey). DECaLS, BASS, and MzLS include data obtained, respectively, at the Blanco telescope at Cerro Tololo Inter-American Observatory and NSF’s NOIRLab; the Bok telescope at Steward Observatory, University of Arizona; and the Mayall telescope at Kitt Peak National Observatory and NOIRLab. Pipeline processing and analyses of the data were supported by NOIRLab and the Lawrence Berkeley National Laboratory (LBNL). The Legacy Surveys project is honored to be permitted to conduct astronomical research on Iolkam Du’ag (Kitt Peak), a mountain with particular significance to the Tohono O’odham Nation. 
See \url{https://www.legacysurvey.org/acknowledgment/} for the full Legacy Surveys acknowledgment.

The GALEX data presented in this paper were obtained from the Mikulski Archive for Space Telescopes (MAST) at the Space Telescope Science Institute (STScI). STScI is operated by the Association of Universities for Research in Astronomy, Inc., under NASA contract NAS5-26555. Support for MAST is provided by the NASA Office of Space Science through grant NAG5-7584 and by other grants and contracts.
All the {\it GALEX} data used in this paper can be found in MAST: \dataset[10.17909/w2jt-c397]{https://doi.org/10.17909/w2jt-c397}.

This work also makes use of data from the Arecibo Legacy Fast ALFA (ALFALFA) survey. ALFALFA is a collaboration between Cornell University and the National Astronomy and Ionosphere Center, with participation by the National Optical Astronomy Observatory and other institutions. The ALFALFA survey has been supported by the National Science Foundation under grants AST-0607007 and AST-1107390, and by the Brinson Foundation.

%\end{acknowledgments}

\bibliography{ref}{}
\bibliographystyle{aasjournal}

\end{document}